\documentclass[referee,sn-basic,Numbered]{sn-jnl}

\usepackage{graphicx}
\usepackage{multirow}
\usepackage{amsmath,amssymb,amsfonts}
\usepackage{amsthm}
\usepackage{mathrsfs}
\usepackage[title]{appendix}
\usepackage{xcolor}
\usepackage{textcomp}
\usepackage{manyfoot}
\usepackage{booktabs}
\usepackage{algorithm}
\usepackage{algorithmicx}
\usepackage{algpseudocode}
\usepackage{listings}
\usepackage{float}
\usepackage{hyperref} 
\usepackage{lineno}
\usepackage{tcolorbox}
\usepackage{soul} 
\usepackage{xcolor} 
\usepackage{booktabs}
\usepackage{makecell}
\usepackage{multirow}
\usepackage{threeparttable}
\usepackage{siunitx}

\definecolor{lightyellow}{rgb}{1, 1, 0.85}

\begin{document}

\title{Leveraging Large Language Models for Career Mobility Analysis: A Study of Gender, Race, and Job Change Using U.S. Online Resume Profiles}

\author*[1]{\fnm{Palakorn} \sur{Achananuparp}}\email{palakorna@smu.edu.sg}
\author[2]{\fnm{Ye} \sur{Xu}}\email{yx2625@columbia.edu}
\author[2]{\fnm{Yao} \sur{Lu}}\email{yl2479@columbia.edu}
\author[1]{\fnm{Xavier Jayaraj Siddarth} \sur{Ashok}}\email{xaviera@smu.edu.sg}
\author[1]{\fnm{Ee-Peng} \sur{Lim}}\email{eplim@smu.edu.sg}

\affil[1]{\orgname{Singapore Management University}, \orgaddress{\city{Singapore}, \country{Singapore}}}
\affil[2]{\orgname{Columbia University}, \orgaddress{\city{New York City}, \state{New York}, \country{USA}}}

\abstract{
We present a large-scale analysis of career mobility of college-educated U.S. workers using online resume profiles to investigate how gender, race, and job change options are associated with upward mobility. This study addresses key research questions of how the job changes affect their upward career mobility, and how the outcomes of upward career mobility differ by gender and race. We address data challenges -- such as missing demographic attributes, missing wage data, and noisy occupation labels -- through various data processing and Artificial Intelligence (AI) methods. In particular, we develop a large language models (LLMs) based occupation classification method known as {\em FewSOC} that achieves accuracy significantly higher than the original occupation labels in the resume dataset. Analysis of 228,710 career trajectories reveals that intra-firm occupation change has been found to facilitate upward mobility most strongly, followed by inter-firm occupation change and inter-firm lateral move. Women and Black college graduates experience significantly lower returns from job changes than men and White peers. Multilevel sensitivity analyses confirm that these disparities are robust to cluster-level heterogeneity and reveal additional intersectional patterns.
}


\maketitle


\section{Introduction}
\label{sec:introduction}

Over the past several decades, the U.S. labor market has undergone significant changes, marked by the decline of internal labor markets and the rise of external labor markets. Internal labor markets, once characterized by stable, long-term employment and opportunities for internal mobility within organizations, have diminished due to shifts in organizational structures and employment practices \cite{diprete2002internal,kalleberg2000nonstandard}. In their place, external labor markets have expanded, driven by flexible employment arrangements and increased movement across organizations \cite{bidwell2013employment,farber2008employment}. As a result, employers now rely more heavily on hiring external candidates, creating multiple pathways for career mobility.

At the same time, the educational composition of the U.S. workforce has undergone dramatic changes.  The labor force has become increasingly educated over time. The share of workers with at least a college degree has risen from 26\% in 1991 to nearly 42\% in 2022~\cite{brundage2017educational}. This shift reflects expanded access to higher education and growing demand for high-skilled workers.

Despite these profound changes, existing research on internal and external job mobility often aggregates workers across educational levels, limiting our understanding of how college graduates -- the fastest-growing segment of the workforce -- navigate these evolving mobility pathways. Furthermore, few studies systematically distinguish between different types of job mobility, potentially obscuring important variations in their prevalence and impact. A critical gap remains: What are the primary modes of job mobility for college-educated workers? Which modes of job mobility facilitate upward career advancement? And how do the returns to various modes of job mobility vary by gender and race?

The college expansion has also brought about increased gender and racial/ethnic diversity in the workforce \cite{ortman2009united}. Between 1980 and 2023, the share of women and ethnoracial minority workers in high-skilled workforce has risen dramatically \cite{pew2024gendergains}. However, little is known about how the frequency and outcomes of internal and external job mobility vary by gender and race. On one hand, persistent structural barriers may limit the opportunities and benefits of job mobility for women and minority groups, compounding existing disadvantages. On the other hand, the increasing flexibility of the labor market may offer them new opportunities for upward mobility, particularly through external job changes that allow individuals to bypass entrenched workplace barriers and leverage external opportunities.

A major obstacle to understanding these dynamics is data availability. Addressing these questions requires a sufficiently large and diverse sample of college-educated workers that can be disaggregated by gender and race/ethnicity while still maintaining robust subgroup sizes. However, general surveys often lack the necessary sample size and rarely collect complete employment histories. At best, they may capture information about a respondent’s first job, most recent job, or a limited series of positions. Full employment trajectories are seldom documented.

Our study addresses the above obstacle through the use of an online resume dataset. This data source enables us to measure various types of job mobility -- both internal and external -- across a large sample of college graduates. Crucially, it provides detailed job histories, allowing us to construct longitudinal employment records for individual workers, unlike studies relying on LinkedIn's aggregate audience estimates \cite{haranko2018professional,kashyap2021analysing,berte2023monitoring,kalhor2024gender}. This approach not only offers a clearer picture of job mobility patterns but also provides new insights into how these patterns interact with gender and racial/ethnic identities.

While the online resume dataset contains very detailed job mobility data, two additional challenges must be addressed: {\em gender and race prediction} and {\em occupation label inference}.  As a secondary research dataset, resumes often do not contain all the essential attribute labels. Attributes such as gender and race of workers, as well as occupational labels for jobs are often missing.  However, these attributes are essential for analyzing job mobility patterns. 

Our study therefore introduces machine learning methods to predict workers' genders and races.  We also developed a novel method for inferring occupation labels -- comprising codes and titles from the Standard Occupational Classification (SOC) system -- based on job titles and company names. Online job-related data, including the resume dataset used in this study, are typically labeled with SOC codes and titles by automated algorithms. However, the accuracy of these inferred SOC labels is often unknown. As demonstrated later in this study, the default SOC labels in our dataset achieve only 63\% accuracy, raising significant concerns about their reliability and impact on the validity of the study’s findings. 

Developing robust occupation classification models is however challenging due to the scarcity of high-quality labeled data and the complexity of SOC taxonomies. To address this limitation, we present a novel \underline{few}-shot\footnote{Few-shot (also known as $k$-shot) prompting refers to a technique where a large language model (LLM) is provided with a small number of $k$ examples in the input prompt to guide it in performing a task. This contrasts with zero-shot prompting, where no examples are given, and the LLM relies solely on its pre-existing knowledge to perform a task. In general, few-shot prompting has been shown to significantly improve task performance compared to zero-shot prompting, as it allows the model to better generalize from limited input \cite{brown2020language}.} prompting framework for \underline{SOC} classification titled \textbf{FewSOC}, which leverages the vast knowledge and reasoning capabilities of large language models (LLMs) like OpenAI's GPT models. Our crowdsourced evaluation demonstrates that the SOC labels generated by FewSOC are significantly more accurate than the default labels provided with the dataset. As a result, we use the improved FewSOC-generated SOC labels in our analysis of career mobility, ensuring more reliable and valid findings.

Building on the context of evolving labor market dynamics and the increasing diversity of the workforce, this study utilizes a large dataset of online resume profiles to address the following research questions.

\textbf{RQ1}: How do college-educated workers navigate different job change options to achieve upward career mobility? Specifically, what factors contribute to their success in achieving upward mobility?

\textbf{RQ2}: How do the outcomes of upward career mobility for college-educated workers differ by gender and race?

This article is organized as follows. We begin with a review of relevant research on career mobility and occupation classification. Next, we describe the online resume dataset, including the preprocessing steps and criteria for selecting the study subset. We then introduce FewSOC, our LLM-based occupation classification method, and outline the crowdsourced evaluation of its effectiveness. This is followed by regression analyses addressing our research questions. Finally, we discuss the study’s limitations, propose directions for future work, and conclude with key insights and implications.

\section{Related Work}
\label{sec:reated_work}

\subsection{Determinants of Career Mobility}
\label{subsec:related_work_mobility}

Job mobility is a key pathway to achieving overall career advancement. However, the likelihood of job mobility is not evenly distributed across all groups. Traditionally disadvantaged populations, such as women and minorities, are less likely to experience mobility, partly due to entrenched biases that hinder their promotion opportunities (internal mobility) and compromise their future employment prospects (external mobility). These barriers constrain the career advancement of minority workers both within and across organizations \cite{bertrand2004emily,pager2003mark,pager2009discrimination,waldinger2003other}.

Another factor contributing to these challenges is the segregation of women and minorities within restricted social networks. These networks often provide limited access to mainstream connections that are crucial for obtaining quality job information, securing referrals, and influencing hiring decisions. As a result, their opportunities for external mobility are diminished \cite{kalfa2018social,moss2001stories,petersen2000offering}. Additionally, women and minority workers are disproportionately concentrated in minority-dominated jobs, which typically offer lower pay and fewer opportunities for upward mobility \cite{huffman2004racial}.

Despite well-documented barriers, little is known about how these dynamics affect highly-educated workers. While white-collar minority professionals may have better mobility prospects than their blue-collar counterparts, they often face a gendered or racialized ``glass ceiling'' \cite{cox1990invisible}. However, the advantages of higher education may also mitigate some barriers, leading to different outcomes for this group. Even when women and minorities achieve job mobility, it is unclear if it results in upward mobility. For those starting in disadvantaged positions, mobility can provide an opportunity to move beyond the lower rungs of the occupational hierarchy, but benefits vary. Women, minorities, and lower-skilled workers face greater barriers and often experience diminished returns from job changes \cite{bukodi2010bad}.

Several factors contribute to these disparities. Racial discrimination and segregated social networks, for example, limit the gains minority workers can achieve through job mobility. In external labor markets, hiring and wage-setting discrimination is exacerbated by incomplete information about a worker’s productivity \cite{bidwell2011paying}. Employers may devalue the credentials of female and minority workers and attach biased social meanings to their job mobility \cite{fuller2008job}. As a result, minority workers who pursue new employment opportunities may receive lower wage offers or be funneled into lower-level positions with limited prospects for advancement. Additionally, segregated social networks further exacerbate these disadvantages by restricting access to high-quality job referrals and information \cite{collins1997black,royster2003race,stainback2008social}. Ultimately, job mobility for women and minorities may often result in inefficient churning between roles of similar status and pay, whether within or across organizations.

Despite the importance of these issues, existing research has yet to systematically distinguish between different types of job mobility. This lack of specificity may obscure variations in how certain forms of mobility are experienced and their potential outcomes. Furthermore, most studies focus either on low-wage workers or on the workforce as a whole without accounting for differences in educational attainment. As a result, our understanding of how these dynamics play out for more highly-educated workers remains limited.

\subsection{Predicting Occupations from Textual Data}
\label{subsec:related_work_classification}
Occupation is an important demographic attribute in social and computational social science research \citep{hu2016language,haranko2018professional,kashyap2021analysing,zhao2021exploring,berte2023monitoring,miyazaki2024public,guo2024evolution,kalhor2024gender}. Previous research has explored various methodologies and data sources for extracting and classifying occupations from text. Simple dictionary-based methods \citep{zhao2021exploring,miyazaki2024public,guo2024evolution} were often employed due to the limited performance of traditional machine-learning classifiers \citep{pan2019twitter}. For machine-learning based approaches, most studies employ traditional multi-class classification models, relying on feature engineering and machine-learning algorithms. Common feature engineering techniques include TF-IDF weighted bag-of-words model, doc2vec, and word2vec \cite{varelas2022employing,mukherjee2021determining,boselli2017using}. Supervised, unsupervised, and ensemble learning algorithms, such as SVM, logistic regression, kNN, and random forests, are widely used for classification \cite{varelas2022employing,mukherjee2021determining,russ2016computer,javed2015carotene}. Additionally, more recent approaches leverage language models \citep{li2023llm4jobs,achananuparp2025multi}. For instance, Li et al. \cite{li2023llm4jobs} proposed LLM4Jobs, a zero-shot query and retrieval framework utilizing Vicuna-33b. Achananuparp et al. \cite{achananuparp2025multi} introduced a multi-stage few-short prompting framework for occupation classification, leveraging LLMs for inferring and reranking.

These studies utilize different types of data. Some researchers classify occupations using only job titles \cite{boselli2017using,javed2015carotene}, while others base their classification on entire job postings \cite{li2023llm4jobs,varelas2022employing,mukherjee2021determining}.  Occupation labels are typically drawn from existing standard occupational classification (SOC) taxonomies with varying levels of granularity, such as ISCO \cite{varelas2022employing,boselli2017using}, ESCO \cite{li2023llm4jobs}, O*NET-SOC \cite{javed2015carotene}, and others \cite{russ2023evaluation,mukherjee2021determining,russ2016computer}. For instance, Javed et al. \cite{javed2015carotene} used 23 major occupational classes from O*NET-SOC, while Li et al. \cite{li2023llm4jobs} and Varelas et al. \cite{varelas2022employing} employed up to 1,000 ESCO codes and 230 ISCO codes, respectively. Other studies, like Russ et al. \cite{russ2023evaluation}, use 840 unique 6-digit SOC codes from the U.S. SOC 2010 system, while Mukherjee et al. (2021) focused on the five most common SOC codes in their dataset. Most datasets used in these studies are not publicly available for research, with the exception of Li et al. \cite{li2023llm4jobs}, which released a synthetic dataset generated by LLMs.

Methodologically, our FewSOC framework shares the most similarity with LLM4Jobs \cite{li2023llm4jobs} and TGRE \cite{achananuparp2025multi}. These methods require no training data and employ a multi-stage pipeline. In the first stage, LLMs generate candidate SOC codes for the test samples. In the second stage, a retrieval module in LLM4Jobs and TGRE, or a SOC mapping and matching module in FewSOC, identifies the most relevant SOC codes from the taxonomy. Unlike LLM4Jobs, which uses zero-shot prompting to classify one job post at a time, FewSOC employs few-shot examples to classify job title-company name pairs. Generally, few-shot prompting tends to enhance LLM performance in classification tasks compared to zero-shot prompting. Compared to TGRE, which incorporates LLM-based reranking in a third stage, FewSOC adopts a more streamlined pipeline by omitting the reranking module. In addition, FewSOC efficiently leverages batch prompting to process multiple test samples per batch, thereby reducing token usage and processing costs. This optimization enhances efficiency while introducing only a minor trade-off in classification accuracy, making FewSOC particularly suitable for large-scale datasets involving hundreds of thousands of resume profiles used in this study.

To our knowledge, this work is among the few studies to utilize over 1,000 fine-grained SOC codes (at the 6-digit level or higher) from the O*NET-SOC taxonomy as class labels for multi-class classification \cite{achananuparp2025multi}. Furthermore, we are the first to investigate the feasibility of crowdsourced SOC code annotation at scale, leveraging a large dataset of more than 10K job titles and company names sourced from online resume profiles.

\section{Dataset, Data Preprocessing, and Attribute Extraction}
\label{sec:dataset}

\subsection{The Lightcast Dataset}
\label{subsec:lightcast}
We use resume profiles collected from various online sources by Lightcast\footnote{https://lightcast.io/}, a labor market analytics company formed through the merger of two leading labor market data providers: Emsi and Burning Glass Technologies. The dataset, obtained in October 2022, contains approximately 141.6 million profiles from people across the United States, collected from online platforms such as LinkedIn and other professional networking sites\footnote{https://kb.lightcast.io/en/articles/7153977-global-data-101}. Lightcast applies machine learning algorithms and other data processing methods to standardize, de-duplicate, merge, and normalize the profiles data across all sources. Additionally, the raw data are enriched with relevant job-related information, such as occupation, industry, and educational level, and more.

Each profile is organized into three sections: job history, education history, and skills. For this study, we focus primarily on a relevant subset of fields from the job history and education history sections. The selected job history fields include job title, employer information (company name, city, and state), start and end dates of the job, occupation (SOC title and code from the O*NET-SOC 2019 taxonomy), industry (NAICS6 title and code), and a binary indicator of the current job (1 indicating the current job, and 0 otherwise). Similarly, the selected education history fields include educational degree name (CIP6 title), educational level name (bachelor's, master's, doctorate, etc.), start and end dates of the degree, school name, and school location (city and state).

\subsection{Data Partitioning}
\label{subsec:partitioning}
Given the substantial size of the Lightcast dataset, it is essential to organize the resume profiles into manageable partitions for convenience and ease of use. To achieve this, we employ a hashing-based method that groups profiles based on the hashed value of their unique identifiers. This approach allows us to efficiently split the original dataset into 284 distinct partitions, with each partition containing approximately 458,000 profiles, ensuring a random and uniform distribution of data. These partitions are henceforth referred to by the abbreviation ``LC'' followed by their sequential index numbers, from LC0 to LC283. 

\subsection{Profile Data Cleaning and Selection}
\label{subsub:profile_data_selection}
For this study, we use resume profiles from 20 partitions (LC0 to LC19), comprising approximately 10 million individuals. Given the volume of data and its inherently noisy nature, there may be discrepancies, inaccuracies, or even fictitious profiles. We define the following data filtering criteria to address these issues and ensure data quality as well as identify relevant profiles:

\begin{enumerate}
    \item \textbf{Job Records}: Every job record must include the job title, company name, city, state, country, and start date. Past job records are also required to have end dates. Job records with end dates preceding start dates were excluded.
    \item \textbf{Job Titles}: Job titles are filtered to ensure they represent standard employment positions. Non-occupational job titles, such as ``student,'' ``intern,'' or ``owner,'' as well as those merging multiple roles into single title were excluded from the profiles. The LLM-based method for identifying such job titles is described in Section \ref{subsec:prompting_method}.
    \item \textbf{Education Records}: Each education record must specify the degree obtained, start date, end date, and school name. Only profiles with at least a bachelor's degree (BA+) were included as our work focuses on college-educated workers. The college-educated workers are also well represented in the online resume data compared to less-educated populations.
    \item \textbf{Post-graduation Gap}: Profiles with a post-graduation gap exceeding specific thresholds are excluded to ensure that the first job record accurately reflects the individual's actual first position after graduation. The post-graduation gap is defined as the time between completing a bachelor's degree and starting full-time employment. Notably, 68.73\% of profiles exhibit a gap of one year or more. The thresholds, established at the 75th percentile of the observed data, represent maximum allowable gaps: 3.75 years for bachelor's degree holders, 5.59 years for master's degree holders, and 8.25 years for doctoral degree holders. For instance, a master's degree holder should enter the workforce no later than 5.59 years after obtaining their bachelor's degree.
    \item \textbf{Timeframe}: Only profiles with job records from 1999 to 2022 are considered. This timeframe corresponds with the years of wage data utilized in the study as described with great details in Section~\ref{subsec:attribute}.
\end{enumerate}

After applying these criteria, the resulting set of approximately 1.5M profiles captures the early career histories of individuals with a bachelor's degree or higher, encompassing standard employment from 1999 to 2022.

\subsection{Career Trajectories Construction}
\label{subsec:career_trajectories}

Many resume profiles contain multiple temporally overlapping job records, complicating career mobility analysis and making it challenging to accurately track career progression. To address this, we constructed a linear career trajectory for each profile, consisting of a sequence of job records where each record follows the previous one without any overlapping time periods.  The job records are first sorted by start date followed by end date.  They are then sequentially added to the career trajectory one at a time skipping the ones that overlap the previously added job record as outlined in Algorithm \ref{algo:construct_trajectory}. Following the construction to remove overlapping jobs, the average length of a linear career trajectory is approximately 70\% of the original job history length.

\begin{algorithm}[H]
\caption{Construct Linear Career Trajectory}
\label{algo:construct_trajectory}
\begin{algorithmic}[1]
\Require{Job history $H = \{j_1, j_2, \ldots, j_n\}$ with job records $j_i = (start_i, end_i)$}
\Ensure{Career trajectory $T$}
\Procedure{ConstructTrajectory}{$H$}
    \State Initialize $T \gets \{\}$
    \State Sort $H$ by $(start_i, -end_i)$ for all $i$ and set $T \gets \{j_1\}$
    \For{$k = 2$ to $n$}
        \State $j_m = \text{last job in } T$
        \If{$(start_k \geq end_m)$}
            \State $T \gets T \cup \{j_k\}$
        \EndIf
    \EndFor
    \State \Return{$T$}
\EndProcedure
\end{algorithmic}
\end{algorithm}

Given the career trajectories dataset, we selected all trajectories with a career length of 5 years or more. For those trajectories exceeding 5 years, we truncated them to include only job records that started within the 5-year window. This step standardizes the observation period across trajectories and selects the relevant early career phases for subsequent analysis. The final dataset used in the analysis, referred to as \textit{Career229K}\footnote{Our naming convention combines the dataset type with its size -- for example, Career229K represents a dataset with approximately 229,000 individual career trajectories.}, comprises 228,710 career trajectories.

\subsection{Attribute Value Extraction}
\label{subsec:attribute}

In this section, we outline the systematic process of extracting attribute values, such as gender, race, and occupation, from resume profiles to enhance the richness of the Career229K dataset. These extracted attributes serve as independent variables in subsequent regression analysis, enabling us to examine their association with individuals' upward mobility within their early careers.

\textbf{Gender and Race}: Each profile is categorized as either male or female for gender and classified into one of the following races: White, Black, Asian, or Hispanic. To infer these attributes without identifying the workers, we utilized a separate dataset provided by Lightcast, containing only unique names extracted from the same set of profiles in our dataset. We applied our gender and race classification models to this set of names and shared the prediction results with Lightcast. They then added the inferred gender and race values to the corresponding profiles and then sent the enriched anonymized profiles (without names) back to us. Gender was inferred using a maximum likelihood model implemented with the Python package \textit{chicksexer}\footnote{https://github.com/kensk8er/chicksexer}, achieving a testing accuracy of 93.4\% \cite{krstovski2023computation}. Race was inferred using a logistic regression classifier trained on multiple full-name datasets, yielding a testing accuracy score of 84\%.


\textbf{Educational Attainment}: Educational attainment is determined by the highest educational degree achieved (bachelor's, master's, or doctorate) within each individual's observed career trajectory. Generally, we expect that higher educational attainment is associated with greater upward mobility, because of increased access to higher-paying jobs and career advancement opportunities.

\textbf{Cohort}: We classify individuals into generational cohorts based on their inferred birth year, calculated by subtracting 23 (the average age of graduation) from the year they obtained their Bachelor's degree. To determine generation, we use definitions from the Pew Research Center \cite{pew2015}. For instance, individuals born between 1981 and 1996 are classified as Millennials, while those born between 1965 and 1980 are categorized as Generation X. Belonging to different cohorts can be linked to career mobility in various ways, because of different economic conditions at job market entry and distinct attitudes toward career advancement.

\textbf{Regional Economic Ranking}: We use the decile of the gross domestic product (GDP) of the state where individuals first started their careers as a proxy for local economic opportunity. States with higher GDP typically have larger job markets and a wider range of employment options, providing greater potential for career advancement. To derive the decile, we collected historical regional GDP data from the Bureau of Economic Analysis (BEA) and computed the deciles of real GDP for each state and year. Then, we retrieved the corresponding decile value for each profile based on the employer's state and the start year of the individual's first job in their career trajectory.

\textbf{Occupational Wage}: Each job record is enriched with state-level occupational wage information obtained from historical estimates provided by the Bureau of Labor Statistics (BLS) for the years 1999 to 2022. We began by compiling all available regional wage data from wage tables on the BLS website, which offered data from 1997 onward at the time of this study.  Two major challenges arose: incompatible SOC codes in older BLS wage tables and the need to standardize the SOC codes across the tables. To resolve these issues, we excluded any wage tables predating 1999 and standardized the SOC codes in the wage tables by mapping the older 6-digit SOC codes to the 2019 version using O*NET-SOC crosswalk tables\footnote{https://www.onetcenter.org/taxonomy.html}. Furthermore, we excluded incompatible SOC codes between the 2019 and 2020 wage tables. For each job record, we retrieved the mean annual wage from the wage tables based on the corresponding year, state, and 6-digit SOC code. In cases where multiple possible SOC matches were possible, we computed the average wage across all applicable SOC codes. Any missing occupational wage data for specific years and states were imputed using linear interpolation.

\textbf{Job Mobility}: Job mobility is broadly defined as movement from one job to another regardless of the nature of the moves. In our study, it is operationalized as the frequency of job changes within the observed career trajectory. Our data indicates that some individuals may exhibit an unusually high number of job changes. To mitigate the influence of these outliers, we top-code the job change frequency at the 95th percentile, capping it at a maximum of four jobs. Generally, job mobility is hypothesized to be positively associated with upward mobility as frequent job changes can lead to opportunities for higher-paying positions and skill development.

\textbf{Job Change Types}: To characterize the nature of job transitions within a career trajectory, we define four categories of job change based on change in both firm affiliation and occupation (as indicated by SOC codes):

\begin{itemize} 
    \item \textbf{Inter-firm Occupation Change (Type 1)}: The trajectory involves the worker making some transition to a role in a different occupation at a different firm. Generally, this type of change is positively associated with the highest potential for upward mobility. 
    \item \textbf{Intra-firm Occupation Change (Type 2)}: The trajectory sees the worker transiting to a role in a different occupation within the same firm. 
    \item \textbf{Inter-firm Lateral Move (Type 3)}:  The trajectory consists of transitions of the worker to a role in the same occupation at a different firm. 
    \item \textbf{Intra-firm Lateral Move (Type 4)}:  The trajectory involves the worker taking a role in the same occupation within the same firm. This includes cases of occupational and firm stability, where workers remain in the same firm and position. Generally, this type of change is least likely to facilitate upward mobility, as it largely reflects continuity.  
\end{itemize}

To capture these job changes, we assign four non-mutually exclusive dichotomous variables to entire career trajectories, indicating whether each category of job change is present in an individual's trajectory. Operationally, these variables are determined by comparing 8-digit SOC codes and company names between job records. For example, if a user has made at least one Type-1 and one Type-2 job change but has never made any Type-3 or Type-4 changes in their career trajectories, their Type-1, Type-2, Type-3, and Type-4 indicators would be 1, 1, 0, and 0, respectively.

\textbf{Occupation and Industry}: Using its proprietary algorithms, Lightcast provides occupation and industry information for each job record using 8-digit SOC codes from the O*NET-SOC 2019 taxonomy and 6-digit industry codes from the North American Industry Classification System (NAICS). These codes are inferred from the context of job records, including job titles and company names. SOC codes not only serve as categorical descriptors for job records but are also integral to various data processing steps in this study, including deriving occupational wages and determining the nature of job changes. Therefore, ensuring the accuracy of the SOC codes is crucial. A manual annotation conducted by the first author of this study on 1,240 job records associated with the 124 most popular SOC codes revealed that Lightcast's assigned SOC codes were approximately 63\% accurate at the 8-digit level, raising significant internal validity concerns. Incorrect SOC assignments can distort career trajectories and lead to erroneous assessments of upward mobility, undermining the reliability of our findings. Consequently, we propose a method to enhance SOC classification using large language models (LLMs) in Section \ref{sec:occupation_classification}.

\section{Occupation Classification using Large Language Models}
\label{sec:occupation_classification}
Fine-grained occupation classification, encompassing over a thousand unique SOC codes and titles, is limited by the scarcity of open-source labeled training data. Without sufficient labeled examples, traditional supervised learning methods often struggle to achieve the accuracy required for reliable career mobility analysis. Meanwhile, large language models (LLMs) have exhibited strong capabilities across a variety of natural language processing (NLP) tasks, due in part to their in-context learning (ICL) abilities \cite{brown2020language}. This allows LLMs to adapt to new tasks with only a few examples, making them a promising option for addressing these data-scarce challenges.

\subsection{FewSOC: Few-shot Batch Prompting Framework for SOC Classification}
\label{subsec:prompting_method}
We present \textbf{FewSOC}, our prompting framework for multi-class classification of SOC codes using large language models (LLMs), which leverages the extensive knowledge of jobs and occupations encoded within LLMs. The method consists of two steps: (1) initial SOC generation using job title and company name as input and (2) SOC mapping and matching using the initial SOC.

In the initial SOC generation step, a SOC label -- comprising an occupational title and an 8-digit code from the O*NET-SOC 2019 taxonomy -- is generated for a given job title-company name pair by an LLM via few-shot prompting. Additionally, the LLM is instructed to perform two auxiliary binary prediction tasks within the same prompt: (a) detecting mentions of non-occupational roles, such as student, intern, etc., and (b) detecting mentions of multiple occupational roles in a job record for the purpose of profile filtering. 

To minimize token usage and processing time when handling millions of job records from the Lightcast dataset, we generate responses for multiple job records per inference batch, similar to the batch prompting approach \cite{cheng2023batch}. The prompt template is shown in \ref{sec:prompt_template}.

Since LLMs have limited precision in generating exact codes from large taxonomies \cite{soroush2024large,lee2024large}, the SOC mapping and matching step addresses any invalid SOC codes produced in the initial SOC generation step -- those not found in the O*NET-SOC 2019 taxonomy. Invalid SOC codes may arise from SOC codes belonging to older taxonomy versions or from non-existent SOC codes. For each invalid SOC code, we iteratively look up O*NET-SOC crosswalk tables to map older SOC versions to the 2019 version. If the process results in multiple possible matches (1-to-many mappings), we then prompt the LLM to perform a batch selection of the most likely SOC code. The prompt template can be found in \ref{sec:prompt_template}. Finally, for each remaining invalid SOC code not found in the crosswalk tables, we measure its similarity to all 1,016 8-digit SOC codes in the O*NET-SOC 2019 taxonomy based on the proportion of word overlap between their titles (in Algorithm \ref{algo:word_overlap}), replacing it with the closest match.

\begin{algorithm}
\caption{Word Overlap Similarity}
\label{algo:word_overlap}
\begin{algorithmic}[1]
\Require $s$: title of the queried SOC code
\Require $t$: title of the target SOC code
\Ensure Proportion of word overlap between $s$ and $t$
\Procedure{word\_overlap\_sim}{s, t}
    \State Convert $s$ and $t$ to lowercase
    \State Split $s$ and $t$ into an array of words
    \State $overlap \gets$ number of words in $s$ that are also in $t$
    \State $length \gets$ number of words in $s$
    \State \Return $overlap / length$
\EndProcedure
\end{algorithmic}
\end{algorithm}

\subsection{Experimental Setup}
\label{subsec:classification_setup}
To evaluate the accuracy of FewSOC, we first constructed a test set of representative occupations by sampling 11,920 job records -- comprising job title-company name pairs -- from the job histories of all profiles in the LC0 partition using the following steps:

\begin{enumerate}
    \item We excluded job titles that contained non-alphanumeric characters or non-ASCII characters.
    \item We excluded job titles with more than seven words.
    \item We selected 6-digit SOC codes provided by Lightcast that were associated with at least 20 job records, resulting in 596 qualifying codes.
    \item We randomly sampled 20 job records for each of the selected SOC codes, yielding 11,920 job records.
\end{enumerate}

Note that the 596 SOC codes selected in step 3 may not necessarily reflect the true occupational distribution of the 11,920 job records, as these SOC codes may not be highly accurate at the 6-digit level. Instead, they were used as a preliminary signal of most commonly represented occupations. We refer to this test set as Jobs12K.

Following the construction of Jobs12K, we performed SOC classification on the entire set using our FewSOC method. A fixed set of k-shot examples (k=17), shown in \ref{prompt:examples}, was manually created. First, we selected a distinct set of 8-digit SOC codes from the O*NET-SOC 2019 taxonomy. Next, we created job title-company pairs for the selected SOC codes by randomly pairing relevant job titles with real and fictional companies and organizations, ensuring that the resulting examples did not appear in the test set. To achieve a balance between efficiency and performance, we included five test samples per inference batch. The initial SOC generation step was conducted via the OpenAI API in August 2023, using the GPT-3.5 Turbo model\footnote{In addition to GPT-3.5 Turbo, We also conducted a separate benchmarking experiment in October 2025 to evaluate the performance of recent efficient LLMs using the exact same FewSOC methodology. The results of this experiment are presented in the Appendix.} (with a 4,096-token context window), selected for its cost-effectiveness, at a temperature setting of zero to encourage deterministic outputs. The total API usage cost was \$1.6 USD. We refer to the final set of SOC codes produced by FewSOC simply as the FewSOC set.

\subsection{Crowdsourced Evaluation}
\label{subsec:amt_evaluation}

\subsubsection{Workers and Task Design}
\label{subsub:amt_workers_design}
To compare the occupation codes provided by Lightcast (LC-SOC) and generated by FewSOC, we conducted a crowdsourced evaluation using Amazon Mechanical Turk (AMT). We selected 27 qualified AMT workers based on the following criteria: (1) holding the Mechanical Turk Masters qualification, granted to those who have consistently submitted high-quality results across a variety of Human Intelligence Tasks (HITs); (2) maintaining at least 95\% HITs approval rate from all their past HITs; and (3) achieving a score of 80 or higher out of 100 on our qualification test, which consisted of five multiple-choice questions assessing the AMT workers' basic understanding of job titles and occupations within the O*NET-SOC taxonomy.

\begin{figure}[h]
    \centering
    \includegraphics[width=\textwidth]{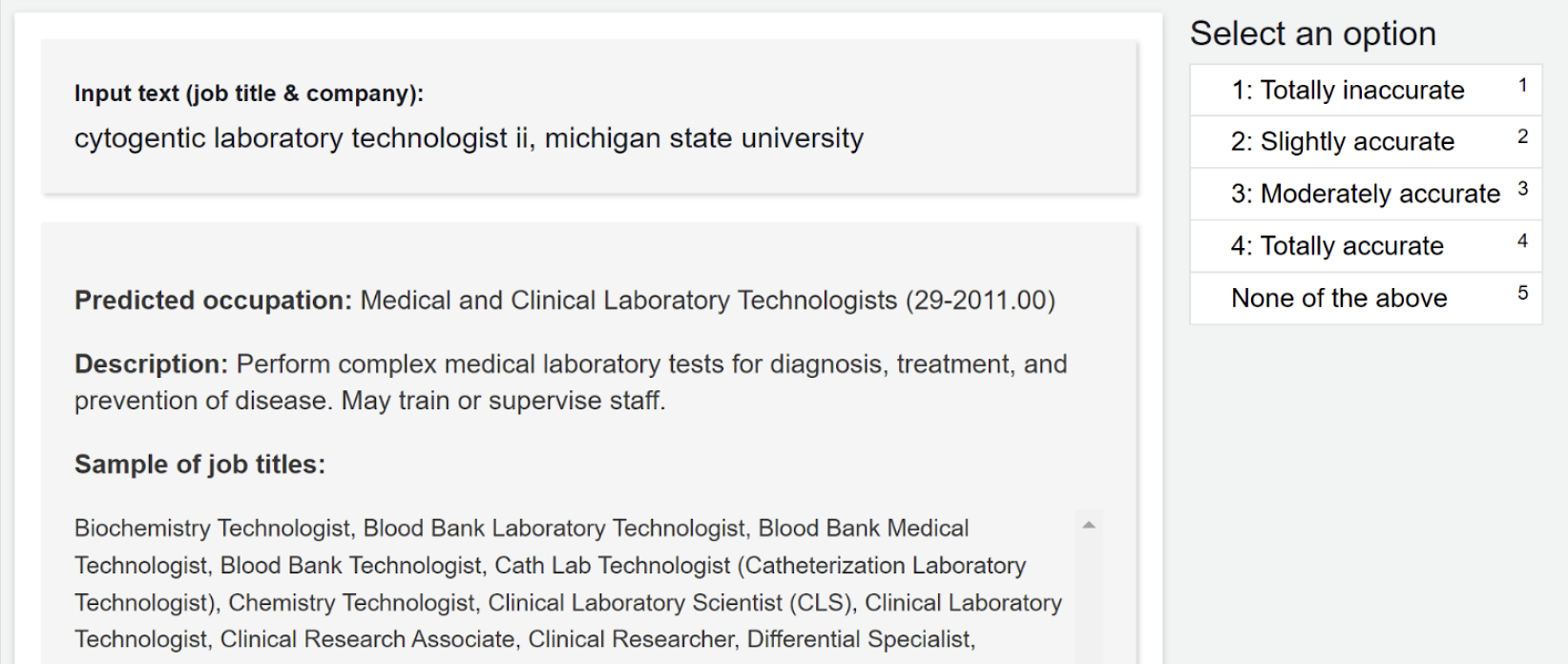}
    \caption{Screenshot of an occupation validation HIT}
    \label{fig:screenshot_hit}
\end{figure}

We designed our crowdsourcing tasks as occupation validation HITs, illustrated in Figure \ref{fig:screenshot_hit}. For each HIT, workers were shown a job title-company name pair as input, an SOC title and 8-digit predicted SOC code (from either LC-SOC or FewSOC) as the predicted occupation, and the corresponding SOC description along with sample job titles from the O*NET-SOC 2019 taxonomy. Workers were instructed to evaluate the accuracy of the predicted occupation based on the input text, selecting a rating from 1 (totally inaccurate) to 4 (totally accurate) using the relevant information provided. Specifically, the following criteria were considered in their assessments:

\begin{itemize}
    \item Skills and knowledge: Predicted occupations sharing similar skills and knowledge with the job in the input text should receive a higher rating.
    \item Job levels: Predicted occupations at a similar job level to that of the input job should receive a higher accuracy rating. For simplicity, we focused on two major job levels: (1) entry-level or experienced positions, and (2) management-level positions. 
\end{itemize}

In addition, we provided a detailed definition and examples for each rating category, emphasizing the distinction between ratings 2 (slightly accurate) and 3 (moderately accurate). Particularly, workers were instructed to assign rating 3 if the predicted occupation describes a related role that employs similar skills and knowledge and is at the same job level or seniority. On the other hand, rating 2 should be assigned when the predicted occupation describes a related role with similar skills and knowledge but at different job levels or seniority.

\subsubsection{Input Data}
\label{subsub:amt_input_data}
To prepare input data for the occupation validation HITs from the Jobs12K set along with the corresponding LC-SOC and FewSOC codes, we first excluded job records containing the keywords, such as student, intern, or owner, as these were easily identifiable as non-occupational titles. We then divided the remaining records into two disjointed sets:
\begin{itemize}
    \item Concordant set: Job records where the LC-SOC and FewSOC codes matched, accounting for approximately 40\% of the data.
    \item Discordant set: Job records where the LC-SOC and FewSOC codes differed, making up the remaining 60\%.
\end{itemize}
The final set of input data for the occupation validation HITs included all job titles-company name pairs and associated SOC codes from the concordant set, which allows us to evaluate cases where LC-SOC and FewSOC agree. Additionally, we included two subsets from the discordant set: one containing LC-SOC codes and the other containing FewSOC codes, enabling us to compare divergent cases. By structuring the input data this way, we effectively reduced the total number of HITs needed, lowering the  cost of the crowdsourcing tasks. This selection resulted in a total of 16,332 HITs for AMT workers to complete.

\subsubsection{Worker Accuracy and Inter-rater Reliability}
\label{subsub:worker_accuracy}
The study was approved by the Institutional Review Board (IRB) of Singapore Management University under Category 1: Exempt from Further IRB Review (approval number IRB-23-142-E039(923)). The occupation validation HITs were conducted between October 2 and October 4, 2023. Each HIT was assigned to three workers and took approximately 10 seconds on average to complete. For quality check, we closely monitored the progress of the HITs and manually validated a small samples of up to 100 responses from each worker, assigning an accuracy score to each. Workers were compensated \$0.03 USD for each approved HIT.

Most workers submitted accurate responses, with an average accuracy score of 0.83 (SD = 0.13). Seventy-two percent of HITs included at least one worker who achieved an accuracy score of 0.8 or higher. However, inter-rater reliability, measured by Krippendorff's alpha ($\alpha$), was relatively low ($\alpha$ = 0.093). Reliability was higher among workers with high accuracy ($\alpha$ = 0.21) and in cases with obvious ratings of 4 (totally accurate) and 1 (totally inaccurate), where $\alpha$ reached 0.52. Disagreements were most frequent when workers rated predicted occupations as 3 (moderately accurate) or 2 (slightly accurate), requiring more nuance. This suggests that making fine-grained occupation judgments at the 8-digit SOC code level is very challenging for non-experts.

\subsubsection{Performance Evaluation of LC-SOC and FewSOC Codes}
\label{subsub:amt_results}
There were 659 and 737 unique 8-digit SOC codes in the LC-SOC and FewSOC codes, respectively. To evaluate the quality of LC-SOC and FewSOC codes, we began by calculating a skill-weighted aggregate rating for each job record and associated SOC by taking a mean rating weighted by worker accuracy to reflect the relative reliability of each worker's responses. Among the discordant set, FewSOC codes achieved a higher aggregate rating (mean = 3.17, SD = 0.66) than LC-SOC codes (mean = 2.95, SD = 0.74), indicating that workers generally found FewSOC codes to be more accurate in case of disagreement. Next, we binarized the aggregate ratings by applying thresholds of 3, signifying a reasonably confident classification. SOC codes with ratings of 3 or greater were considered correct and those below as incorrect. Finally, we computed a precision score for both FewSOC and LC-SOC codes, where precision is defined as the number of job records with correct codes divided by the total number of job records. FewSOC achieved precision scores of 0.72, which is significantly higher by 10.42\% ($p < 0.01$; z-statistic = 10.85) than LC-SOC's precision score of 0.65. These results indicate that our FewSOC framework is more effective at SOC classification than Lightcast's approach. 

Given the superior performance of FewSOC in SOC classification, we chose to utilize the framework to classify SOC codes for resume profiles in the Career229K dataset used in the study.

\section{Effects of Gender, Race, and Job Change on Career Mobility}
\label{sec:method}

With the accurately inferred SOC codes using FewSOC, we gather the career trajectories and demographic attributes of individuals from the Career229K dataset and model their career mobility using logistic regression models as described in Sections~\ref{subsec:career_data} and \ref{subsec:model} respectively. The model results are subsequently discussed in Section~\ref{sec:results}

\subsection{Career Trajectories Data}
\label{subsec:career_data}
The main outcome variable of the analysis is on \textit{upward mobility}, defined as a binary indicator that denotes whether an individual has transitioned into a position with a higher wage by the fifth year of their observed career trajectory. Specifically, let $upward_i$ denotes the upward mobility outcome of individual $i$, and let $w_{i,1}$ and $w_{i,5}$ denote the wages of individual $i$ at the start of their career (first year) and in their fifth year, respectively. If $w_{i,5} - w_{i,1} > 0$ then $upward_i$ = 1; otherwise, $upward_i$ = 0.

The data used in the analysis includes 228,710 career trajectories from the Career229K dataset, as described in Section \ref{subsec:career_trajectories}. Each trajectory contains demographic attributes such as gender, race, educational attainment, and social generation, along with key career-related attributes about the individual's first job. These include the SOC code, industry, occupational wage (i.e., individuals with the same SOC code share the same wage), and the regional economic ranking at the start of their careers. Additionally, it incorporates four binary attributes representing different types of job changes within the career trajectory. The process for extracting these attributes from the resume profiles is described in Section \ref{subsec:attribute}.

Among the individuals (college graduates) in the dataset, 53.71\% achieved upward mobility within the first 5 years of their career. The gender distribution was balanced, with 50.52\% men and 49.48\% women. The racial composition was as follows: 64.89\% White, 13.99\% Black, 12.99\% Asian, and 8.12\% Hispanic. See Table~\ref{tab:demo_dist} for full demographic breakdowns. 

\begin{table}[ht]
\caption{Distribution of demographic attributes in the Career229K dataset}
\centering
\begin{tabular}{@{}llrr@{}}
\toprule
Variable & Category & \multicolumn{1}{l}{Count} & \multicolumn{1}{l}{\%} \\ \midrule
\multirow{2}{*}{Gender} & Male & 115,555 & 50.52 \\
 & Female & 113,155 & 49.48 \\ \midrule
\multirow{4}{*}{Race} & White & 148,418 & 64.89 \\
 & Black & 32,003 & 13.99 \\
 & Asian & 29,718 & 12.99 \\
 & Hispanic & 18,571 & 8.12 \\ \midrule
\multirow{3}{*}{Educational Attainment} & Bachelor's Degree & 168,429 & 73.64 \\
 & Master's Degree & 48,882 & 21.37 \\
 & Doctorate & 11,399 & 4.98 \\ \midrule
\multirow{2}{*}{Social Generation} & Millennials & 179,674 & 78.56 \\
 & Generation X & 49,036 & 21.44 \\
 \bottomrule
\end{tabular}
\label{tab:demo_dist}
\end{table}

Table~\ref{tab:job_change_type_dist} provides the distribution of different job changes types in the dataset. For each type, each count represents the number of users who have experienced at least one instance of the corresponding job change type in their employment history. The majority of individuals (52.86\%) experienced inter-firm occupation changes (Type-1) at least once. Intra-firm occupation changes (Type-2) accounted for 26.29\% of individuals, while inter-firm lateral moves (Type-3) represented 20.12\% of individuals. Lastly, 24.04\% of individuals experienced at least one instance of intra-firm lateral moves (Type-4). 

\begin{table}[ht]
\caption{Distribution of job change types in the Career229K dataset}
\centering
\begin{tabular}{@{}lrr@{}}
\toprule
\multicolumn{1}{l}{Job Change Type} & \multicolumn{1}{l}{Count} & \multicolumn{1}{l}{\%} \\ \midrule
Inter-firm Occupation Change (Type 1) & 120,899 & 52.86 \\
Intra-firm Occupation Change (Type 2) & 60,138 & 26.29 \\
Inter-firm Lateral Move (Type 3) & 46,006 & 20.12 \\
Intra-firm Lateral Move (Type 4) & 54,991 & 24.04 \\ \bottomrule
\end{tabular}
\footnotetext{Each count represents the number of individuals who have experienced at least one instance of each job change type in their career trajectory.}
\label{tab:job_change_type_dist}
\end{table}

\begin{figure}[ht]
\centering
\includegraphics[width=0.8\textwidth]{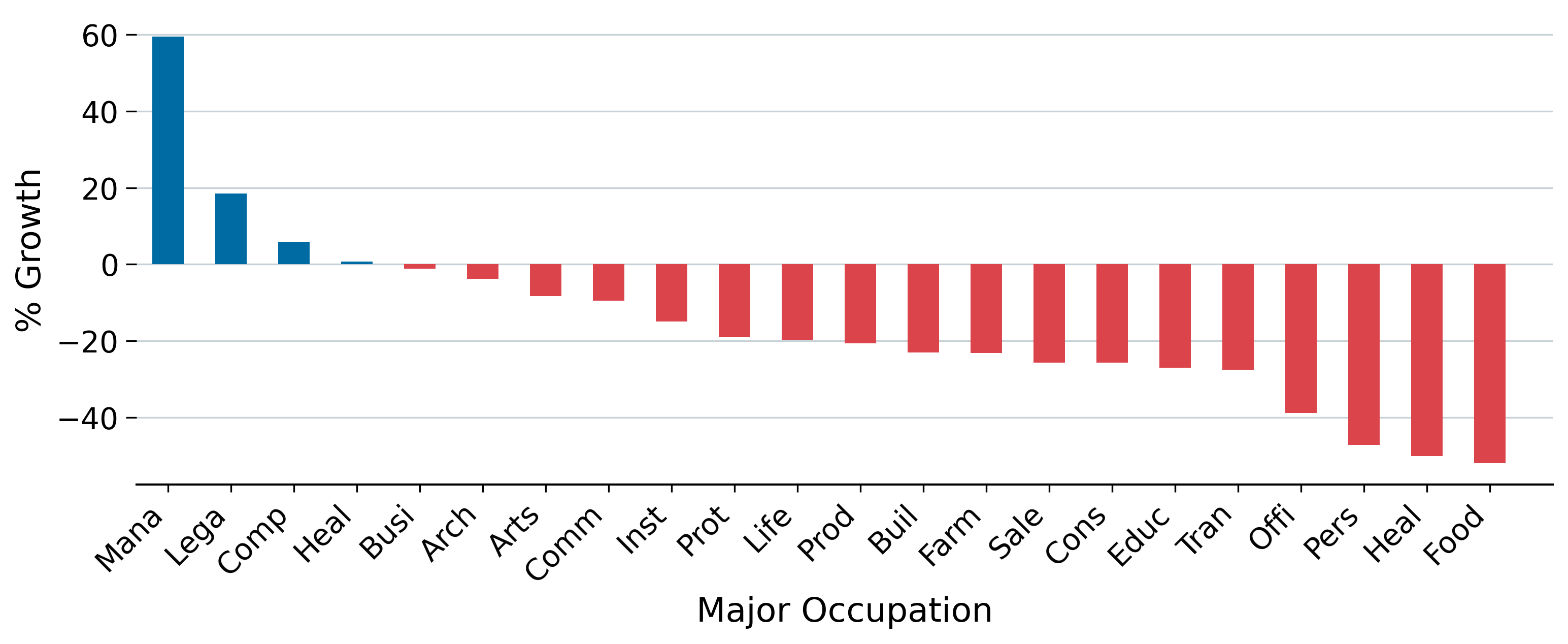}
\caption{Five-year growth rate of each major occupation}
\label{fig:occupation_growth}
\end{figure}

Next, the majority of individuals began their careers in Management, Business and Financial Operations, Sales and Related, or Computer and Mathematical occupations, collectively representing 53.31\% of all profiles, as shown in Table~\ref{tab:occupation_dist} and Figure~\ref{fig:occupation_growth}. After five years, the top three occupations with the highest growth rates were Management, Legal, and Computer and Mathematical. In contrast, the bottom three occupations with the largest negative growth rates were Food Preparation and Serving Related, Healthcare Support, and Personal Care and Service. This follows the intuition that most people tend to transition into higher-wage occupations while moving away from lower-wage ones over time. 

\begin{table}[ht]
\caption{Distribution of occupations in the Career229K dataset}
\centering
\begin{tabular}{@{}lrrrrr@{}}
\toprule
\multirow{2}{*}{\textbf{SOC Title (2-digit Code)}} & \multicolumn{2}{c}{\textbf{Year 1}} & \multicolumn{2}{c}{\textbf{Year 5}} \\ 
& \textbf{Count} & \textbf{\%} & \textbf{Count} & \textbf{\%} \\
\midrule
Management (11) & 41,511 & 18.15 & 66,241 & 28.96 \\
Legal (23) & 4,843 & 2.12 & 5,736 & 2.51 \\
Computer and Mathematical (15) & 20,316 & 8.88 & 21,524 & 9.41 \\
Healthcare Practitioners and Technical (29) & 10,084 & 4.41 & 10,158 & 4.44 \\
Business and Financial Operations (13) & 37,148 & 16.24 & 36,752 & 16.07 \\
Architecture and Engineering (17) & 14,119 & 6.17 & 13,587 & 5.94 \\
Arts, Design, Entertainment, Sports, and Media (27) & 12,573 & 5.50 & 11,526 & 5.04 \\
Community and Social Service (21) & 6,198 & 2.71 & 5,612 & 2.45 \\
Installation, Maintenance, and Repair (49) & 982 & 0.43 & 835 & 0.37 \\
Protective Service (33) & 1,404 & 0.61 & 1,136 & 0.50 \\
Life, Physical, and Social Science (19) & 10,502 & 4.59 & 8,428 & 3.69 \\
Production (51) & 1,686 & 0.74 & 1,339 & 0.59 \\
Building and Grounds Cleaning and Maintenance (37) & 299 & 0.13 & 230 & 0.10 \\
Farming, Fishing, and Forestry (45) & 121 & 0.05 & 93 & 0.04 \\
Sales and Related (41) & 22,956 & 10.04 & 17,055 & 7.46 \\
Construction and Extraction (47) & 583 & 0.25 & 433 & 0.19 \\
Educational Instruction and Library (25) & 16,894 & 7.39 & 12,321 & 5.39 \\
Transportation and Material Moving (53) & 1,418 & 0.62 & 1,028 & 0.45 \\
Office and Administrative Support (43) & 18,977 & 8.30 & 11,600 & 5.07 \\
Personal Care and Service (39) & 2,581 & 1.13 & 1,362 & 0.60 \\
Healthcare Support (31) & 1,266 & 0.55 & 632 & 0.28 \\
Food Preparation and Serving Related (35) & 2,249 & 0.98 & 1,082 & 0.47 \\ \bottomrule
\end{tabular}
\footnotetext{Distribution of occupations by SOC titles and 2-digit codes, ranked from highest to lowest growth rates between Year 1 and Year 5.}
\label{tab:occupation_dist}
\end{table}

\begin{figure}[ht]
\centering
\includegraphics[width=0.85\textwidth]{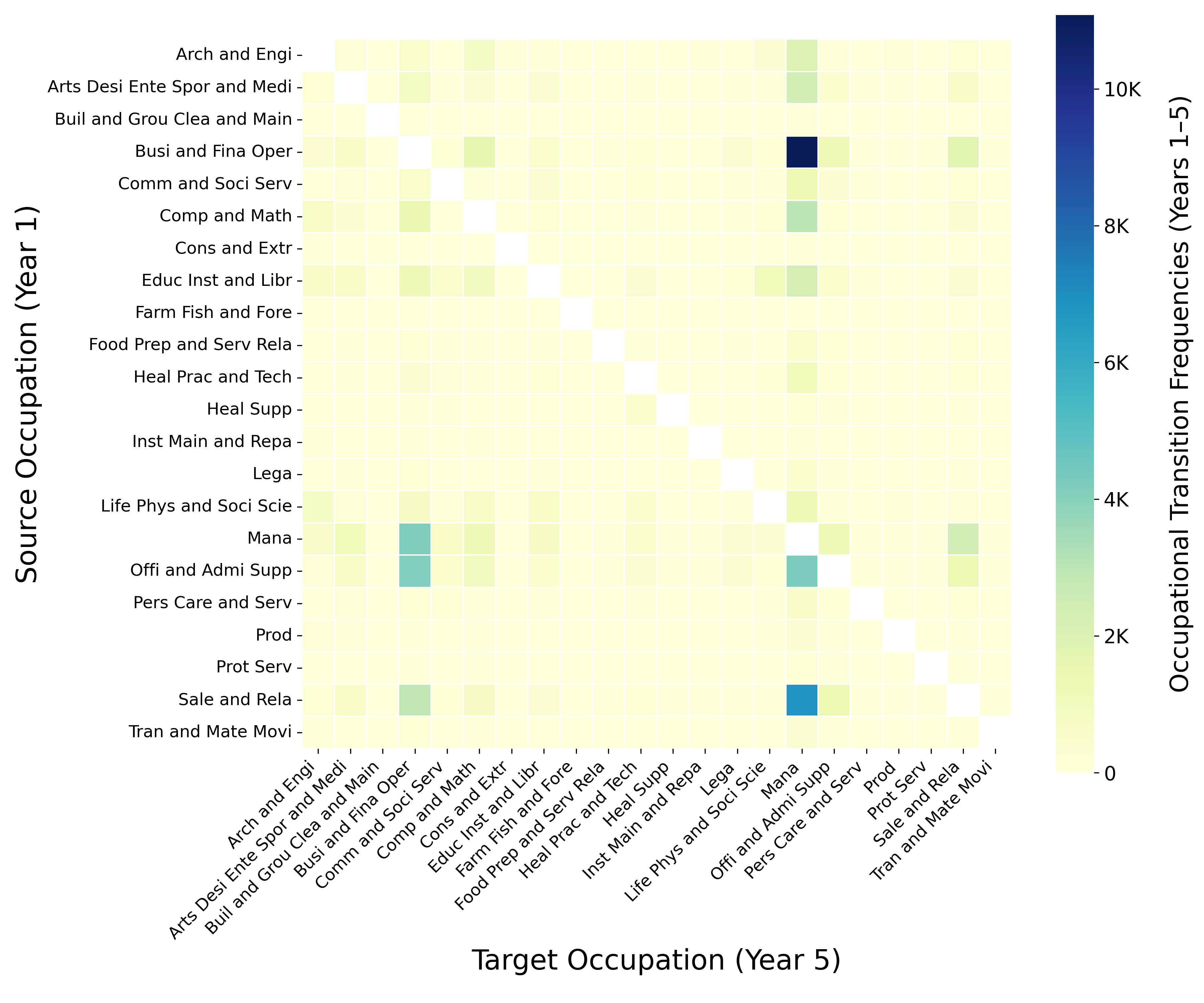}
\caption{Major occupational transitions from year 1 to year 5. Occupation labels are truncated to four characters, e.g., ``Mana'' for Management. Self-transitions (stationary transitions) are excluded.}
\label{fig:occupation_transitions}
\end{figure}

Figure~\ref{fig:occupation_transitions} presents a heatmap visualizing major occupational transitions from Year 1 (source occupation) to Year 5 (target occupation). Each cell in the matrix indicates the frequency of transition between the corresponding source and target occupations, with color intensity proportional to the frequency. The diagonal cells represent individuals who remained in the same major occupation over the five-year period. The Management (Mana) occupation exhibits the highest inbound transition frequency, functioning as a central destination hub in the occupational transition network. This pattern is clearly visible in the dark column corresponding to "Mana" as the Target Occupation. Specifically, the three most frequent transitions between major occupations are all inbound to Management, occurring from: Business and Finance Operations to Management, Sales and Related Occupations to Management, and Office and Administrative Support to Management.

\subsection{Career Mobility Modeling}
\label{subsec:model}

To examine factors predictive of individuals' likelihood of achieving upward mobility within the first five years of their career, we fit four logistic regression models to estimate the effects of the independent variables on the dependent variable (upward mobility), considering both main effects and interactions. While our data clearly exhibit a nested structure (e.g., individuals within firms and occupations), we employ standard logistic regression for the main analysis for several methodological and substantive reasons. Importantly, the modeling choice aligns with our research questions, which focus on population-level gender and racial disparities in career mobility rather than cluster-specific effects.

First, logistic regression coefficients correspond to population-averaged effects, which represent weighted averages across heterogeneous groups. These estimates are typically more conservative than cluster-specific effects and are highly relevant for policy analysis, as they capture the average expected impact of interventions (e.g., reducing wage gaps, facilitating job mobility) across the entire population, irrespective of unobserved cluster characteristics. Second, this modeling choice facilitates direct comparison with the extensive body of mobility and labor economics research that relies on standard generalized linear models (GLMs). Finally, the resulting odds ratios offer straightforward interpretation for both scholarly and policy audiences. To ensure that these population-averaged estimates are not biased by unobserved cluster-level heterogeneity, we also provide multilevel sensitivity analyses in the Appendix.

We apply N-1 dummy coding, setting the reference categories for categorical variables. Specifically, we set `Male' as the reference for gender, `White' for race, `Bachelor’s degree' for educational attainment, `Generation X' for social generation, `Management' for occupation, and `Agriculture, Forestry, Fishing, and Hunting' for industry. Crucially, our analyses aim to establish statistical associations and predictive relationships between these variables, not causal effects. The model specifications are as follows.

\textbf{Main-effect Model}: {\bf Model 1} examines how socioeconomic variables, including gender, race, educational attainment, social generation, regional economic ranking, occupational wage (logarithmic scale), job mobility, and the three job change types (Types 1, 2, and 3) predict upward mobility. Type-4 job change is omitted due to its inherent relationship with the other job change types, which could introduce multicollinearity into the model. Additionally, we control for the occupation (2-digit SOC) and industry (2-digit NAICS) categories. 

\textbf{Interaction Models}: We explore how gender (\textbf{Model 2}) and race (\textbf{Model 3}) are associated with upward mobility in relation to all three job change types (Types 1, 2, and 3). In addition, {\bf Model 4} incorporates interaction effects between race and all three job change types, stratified by gender cohort, to capture nuanced differences across both racial and gender groups.

The analysis was conducted using the Python programming language (3.11.5) with the following libraries: pandas (2.0.3), Numpy (1.24.3) and Statsmodels (0.14.0).

\section{Results and Discussion}
\label{sec:results}

We examine the research questions by presenting and interpreting the regression results outlined in Table \ref{tab:regression_results}. Specifically, we aim to uncover the effects of job change types and demographic factors on upward mobility (RQ1) and explore how gender and race are linked to these outcomes (RQ2).

\begin{figure}[ht]
\centering
\includegraphics[width=0.9\textwidth]{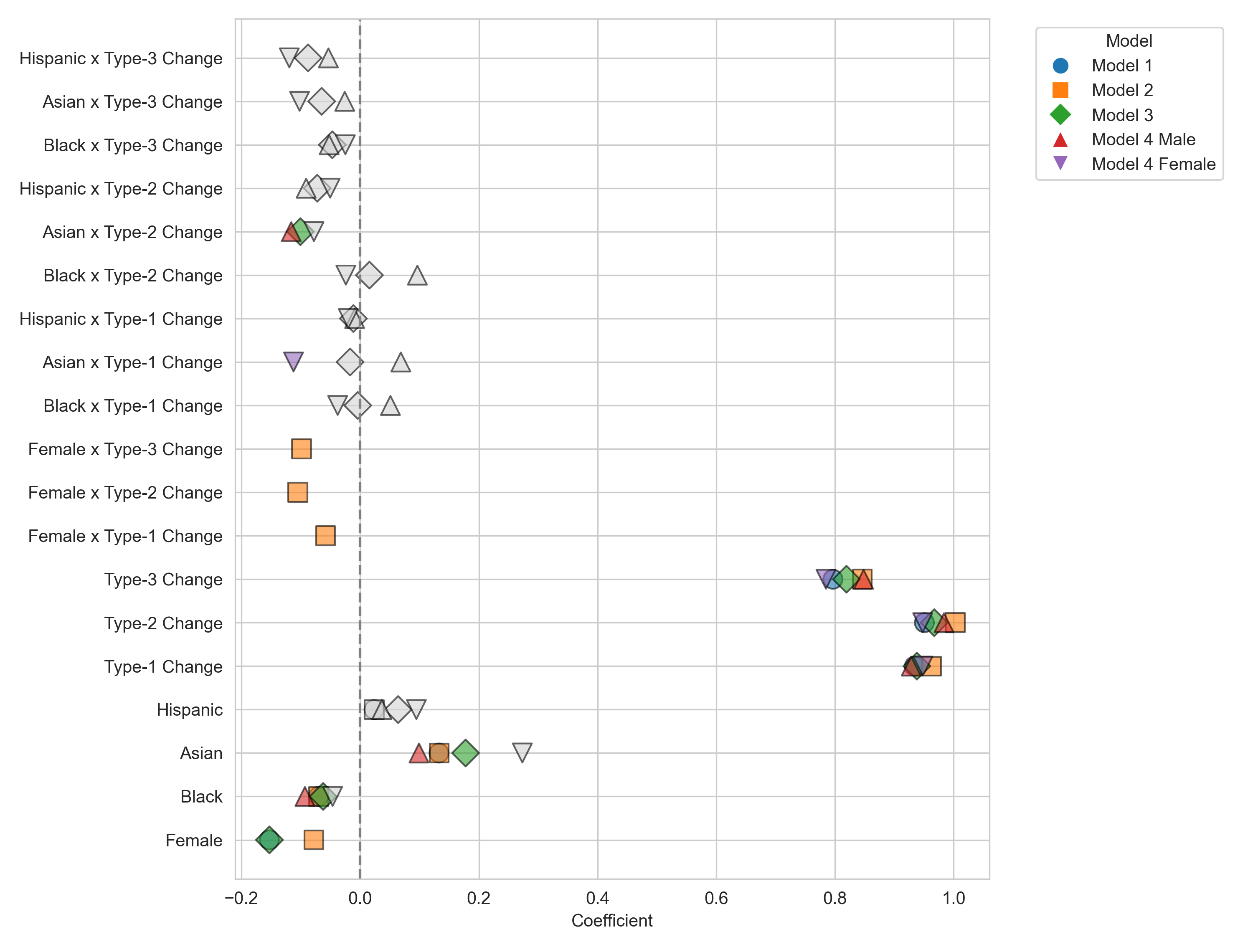}
\caption{Coefficient plot of estimated coefficients for main effects and key interaction terms in Models 1--4. Statistical significance is visually encoded: Coefficients significant at $p<0.05$ are displayed in their full, distinct model colors, while insignificant coefficients ($p>0.05$) are displayed using gray-filled shapes.}
\label{fig:main_effects_plot}
\end{figure}

To facilitate the comparison of estimates across models, Figure \ref{fig:main_effects_plot} displays a coefficient plot for the main effects and key interaction terms. In this figure, statistical significance of each estimate is encoded as follows. Significant coefficients ($p<0.05$) are displayed using the full, distinct model colors. In contrast, insignificant coefficients ($p>0.05$) are displayed using gray-filled shapes. The coefficients for all control variables are presented in Supplementary Figure~\ref{fig:covariates_plot} in the Appendix.

\subsection{RQ1: Effects of Job Change Types and Demographic Factors on Upward Mobility}

The results indicate that all three types of job change -- Type-1, Type-2, and Type-3 changes -- are significantly associated with upward mobility ($p < 0.001$). Among them, Type-2 change (intra-firm occupational change) has the strongest positive effect on achieving upward mobility, followed by Type-1 (inter-firm occupational change), and Type-3 (intra-firm lateral move) changes.

Several demographic factors, including gender, race, and educational attainment, also play a significant role. Specifically, among disadvantaged and minority populations, women face greater challenges in attaining upward mobility than men ($p < 0.001$), and Black college graduates are at a greater disadvantage in achieving upward mobility than their White peers ($p < 0.001$). In contrast, Asian college graduates are more likely to achieve upward mobility than White college graduates ($p < 0.001$). No significant differences are observed between Hispanic and White college graduates in career mobility outcomes. Educational attainment follows a clear trajectory. Specifically, those with doctoral degrees are the most likely to experience upward mobility, followed by those with master’s degrees, while those with bachelor’s degrees have the lowest likelihood among the three groups ($p < 0.001$).

Importantly, these population-averaged effects are robust to unexplained heterogeneity across occupation, industry, state, cohort, and firm size levels. The multilevel sensitivity analysis in the Appendix (Models 1A–4F) shows that the coefficients for gender and race remain directionally consistent and credible across all models. This suggests that the observed disparities are not artifacts of workers being concentrated in particular clusters, instead, they reflect systematic population-level differences.

\begin{table}[!ht]
\caption{Estimated coefficients for main effects and key interaction terms in Models 1--4}
\centering
\begin{tabular}{@{}lccccc@{}}
\toprule
\textbf{Variable} & \textbf{Model 1} & \textbf{Model 2} & \textbf{Model 3} & \multicolumn{2}{c}{\textbf{Model 4}} \\ \cmidrule(l){5-6} 
 & \multicolumn{1}{l}{} & \multicolumn{1}{l}{} & \multicolumn{1}{l}{} & \textbf{Male} & \textbf{Female} \\ \midrule
Intercept & 20.9428* & 20.901* & 20.938* & 20.8066* & 21.1082* \\
Female & -0.153* & -0.0781* & -0.1529* & - & - \\
Black & -0.0701* & -0.0701* & -0.0624$^{\dagger}$ & -0.0931$^{\dagger}$ & -0.0464 \\
Asian & 0.1328* & 0.1332* & 0.1776* & 0.0991$^{\dagger}$ & 0.2732 \\
Hispanic & 0.0234 & 0.0236 & 0.0639 & 0.0365 & 0.0947 \\
Doctorate & 1.2052* & 1.2058* & 1.2045* & 1.1806* & 1.2577* \\
Master's Degree & 0.4082* & 0.4083* & 0.4079* & 0.4046* & 0.4155* \\
Millennials & 0.1739* & 0.1738* & 0.1736* & 0.1831* & 0.1633* \\
Regional Economic Ranking & 0.0217* & 0.0217* & 0.0217* & 0.0149* & 0.0295* \\
Log(Wage) & -2.0908* & -2.0905* & -2.0911* & -2.0749* & -2.1254* \\
Job Mobility & 0.495* & 0.4951* & 0.4948* & 0.5119* & 0.4784* \\
Type-1 Change & 0.9338* & 0.9625* & 0.9379* & 0.9284* & 0.948* \\
Type-2 Change & 0.9506* & 1.0027* & 0.9672* & 0.9836* & 0.9477* \\
Type-3 Change & 0.7965* & 0.8458* & 0.8193* & 0.8481* & 0.7847* \\
Female x Type-1 Change & - & -0.0582$^{\ddagger}$ & - & - & - \\
Female x Type-2 Change & - & -0.1049* & - & - & - \\
Female x Type-3 Change & - & -0.0989* & - & - & - \\
Black x Type-1 Change & - & - & -0.0041 & 0.051 & -0.0379 \\
Asian x Type-1 Change & - & - & -0.017 & 0.0685 & -0.1123$^{\dagger}$ \\
Hispanic x Type-1 Change & - & - & -0.0114 & -0.0093 & -0.02 \\
Black x Type-2 Change & - & - & 0.0156 & 0.0964 & -0.0241 \\
Asian x Type-2 Change & - & - & -0.1007$^{\ddagger}$ & -0.1162$^{\dagger}$ & -0.0779 \\
Hispanic x Type-2 Change & - & - & -0.0724 & -0.091 & -0.0506 \\
Black x Type-3 Change & - & - & -0.0469 & -0.0522 & -0.025 \\
Asian x Type-3 Change & - & - & -0.0649 & -0.026 & -0.1021 \\
Hispanic x Type-3 Change & - & - & -0.0878 & -0.0535 & -0.1194 \\ \bottomrule
\end{tabular}
\footnotetext{Note: \textit{Reference Categories:} Male (Gender), White (Race), Bachelor’s degree (Educational Attainment), Generation X (Social Generation), Management (Occupation), and Agriculture, Forestry, Fishing, and Hunting (Industry). $\dagger$ denotes $p<0.05$, $\ddagger$ denotes $p<0.01$, and * denotes $p<0.001$.}
\label{tab:regression_results}
\end{table}

\subsection{RQ2: Gender and Race Differences in Upward Mobility Outcomes}

The analysis confirms significant gender and race disparities in upward career mobility. Firstly, women are consistently less likely to achieve upward mobility across all job change types ($p < 0.001$). The strong negative interaction effects between being female and Type-2 and Type-3 job changes are especially notable. This indicates that, whenever women change jobs within the same or different firms, they tend to experience lower returns on career mobility than men.

Compared to gender, the interaction effects between race and upward career mobility are more limited. Specifically, no significant interaction effects between race and upward mobility are observed, except for a negative interaction effect between Asian workers and Type-2 job change ($p < 0.01$). The finding suggests that intra-firm transitions tend to be more effective for White than Asian workers.

More nuanced patterns are observed when considering the complex interplay between gender, job change types, and racial differences among male and female cohorts. No significant interaction effects between race and job change types are observed, except for two cases. Firstly, Asian women with Type-1 job changes are significantly at a greater disadvantage in achieving upward mobility than White women ($p < 0.05$). Secondly, Asian men with Type-2 job changes are less likely to experience upward mobility than White men ($p < 0.05$). The results suggest that while different job change types contribute to career mobility, their effects are also conditioned by both gender and race, albeit to a limited extent. 

Notably, these interaction effects are robust when accounting for unobserved heterogeneity via joint random effects for firm size and occupation. The multilevel sensitivity analysis (Models 2F–4F in the Appendix) confirms that the negative Female × Type-2 and Type-3 interactions, as well as the Asian × Type-2 effect for men, remain credible. Additionally, the mixed-effects models reveal new intersectional patterns: Asian men show advantages in Type-1 changes, Black men experience gains in Type-2 changes, Asian women experience disadvantages in Type-1 and Type-3 job changes, and Hispanic women face disadvantages in Type-3 changes. These findings highlight subtle, context-dependent differences in upward mobility that are not fully captured by the population-averaged estimates.

\section{Limitations and Future Work}
\label{sec:limitations}

\textbf{In-context examples}: In this work, we manually engineered a fixed set of in-context examples for FewSOC. While this approach yielded substantial improvements over Lightcast's SOC predictions, it may not guarantee optimal performance and could lead to variability in the quality and representativeness of the examples. Moreover, the manual selection method is time-consuming and inherently subjective. Future research should explore more systematic methods for selecting in-context examples, such as kNN-based approaches \cite{liu2022makes}, to further enhance the effectiveness of FewSOC.

\textbf{Predicted SOC Codes}: The accuracy of predicted SOC codes, while significantly improved by our FewSOC framework compared to Lightcast, remains a potential source of error in this study. Misclassifications of job roles due to prediction errors can bias career trajectory patterns and skew coefficient estimates in regression analyses. The impact of these errors might differ across various occupations, potentially resulting in misleading conclusions for certain groups.

\textbf{Inter-rater reliability}: The low inter-rater agreement among AMT workers raises concerns about the validity of the crowdsourced evaluation of SOC codes. Although we implemented skill-weighted rating aggregation to mitigate this issue, further efforts are needed to improve the reliability and consistency of human annotation in future studies. This could involve providing more comprehensive training for SOC coding, establishing clearer guidelines, or using more advanced techniques for quality control.

\textbf{Gender and race inference}: Our measures of gender and race are derived algorithmically and are therefore less precise than self-reported survey data. This limitation is particularly relevant for race classification, where name-based methods exhibit uneven performance across groups. Inferences for Black and Asian names tend to be less accurate than White and Hispanic names, which introduces a known form of measurement error into the regression models. Econometric theory suggests that such errors in explanatory variables generally dampen estimated effects rather than produce spurious associations. Therefore, the coefficients on race should be interpreted as conservative estimates. Readers should keep these in mind when interpreting the magnitude of the estimated race effects, especially for Black and Asian workers. Future research could enhance the precision of demographic inference by incorporating additional signals (e.g., images) as such data become more widely available in online resume databases.

\textbf{Representation bias}: The study utilizes online resume profiles which provide a sample of online job seekers and may not accurately represent the entire U.S. workforce. Individuals with higher education levels, strong online presences, those with well-established careers, or those actively seeking employment may be overrepresented. Online job searches have played an increasingly central role in how people find work, especially for college graduates, surpassing other means such as personal or professional connections \cite{smith2015}. An estimated 60\% to 70\% of all job vacancies are posted online, with a disproportionate number of them for high-skilled jobs \cite{carnevale2014understanding}.  Almost 80\% of Americans have searched for jobs online in the last two years \cite{smith2015}.  Therefore, we expect the online resume data to represent the highly-educated workforce reasonably well, which is our focus.  Future research could consider addressing potential representation biases from online sources, particularly when studying a broader population.

\textbf{Definition of upward mobility}: We define upward mobility solely on wage increases over five years. This may not fully capture all aspects of career mobility, such as socioeconomic status or occupational privileges. Based on the assumption that there is a strong positive correlation between wage and other socioeconomic factors, we believe wage increase is a reasonable proxy for upward mobility.  Moreover, we use occupational average wage rather than individual wage to assess upward mobility. This underestimates wage variation among individuals within the same occupation. Future research could incorporate more granular firm- or geographic-level data to improve accuracy as such information becomes available. 

\textbf{Scope of analysis}: The study focuses on the first five years of an individual's career. Expanding the analysis to a longer horizon would provide additional insight into long-term career trajectories and upward mobility patterns beyond the initial career stage.

\section{Conclusion}
\label{sec:conslusion}

This study systematically examines how job changes, both within and across firms, shape upward mobility among college graduates in the U.S. and how the returns to job mobility vary by gender and race/ethnicity.
We first addressed major challenges in analyzing career trajectories: the presence of inaccurate occupational labels, missing demographic attributes, and incomplete wage data, through various data processing, integration, and cleaning steps, as well as large language models (LLMs). Specifically, we developed FewSOC, an LLM-based occupation classification framework that improves Standard Occupation Classification (SOC) coding accuracy. A crowdsourced evaluation with qualified Amazon Mechanical Turk (AMT) workers confirmed its accuracy over existing SOC labels, providing more precise career trajectory data for the analysis.

The study advances the existing literature by shifting the focus to an increasingly salient segment of the workforce -- individuals with at least a bachelor's degree -- who navigate a labor market that is often marked by high job mobility and evolving career trajectories. Additionally, our study moves beyond prior research by explicitly differentiating between intra-firm and inter-firm mobility and analyzing them in a unified framework, recognizing that these distinct pathways may yield different career returns. By leveraging online resume data, we construct longitudinal career trajectories over five years for workers from four major ethnoracial groups. Importantly, our approach allows us to distinguish among different types of job transitions.

Our findings reveal that while job changes -- both within and between firms -- are generally associated with upward mobility, the magnitude of these returns varies notably by gender and, to a lesser extent, by race/ethnicity. Women tend to experience lower returns from job changes compared to men, regardless of whether they move within or across firms. This finding highlights persistent gender disparities in career progression. In terms of racial differences, our results indicate that Asian graduates face lower returns to job mobility compared to their White counterparts, while other racial/ethnic groups do not exhibit significant disadvantages in career mobility outcomes. It is essential to note that these results reflect statistical associations and differential career outcomes, not direct causal claims, given the observational nature of the data.

Critically, these disparities are generally robust to unobserved heterogeneity at the occupation and firm-size levels, as shown in the multilevel sensitivity analysis. Furthermore, the mixed-effects models reveal additional complex intersectional patterns between race and mobility: Asian men experience advantages in Type-1 changes, Black men gain from Type-2 changes, Asian women face disadvantages in Type-1 and Type-3 changes, and Hispanic women experience disadvantages in Type-3 changes. These nuanced patterns highlight context-dependent differences in upward mobility that are not fully captured by population-averaged estimates.

By systematically distinguishing between different forms of job mobility and analyzing their differential effects across gender and racial/ethnic groups, our study provides a more comprehensive understanding of labor market inequalities in career advancement among the most educated segment of the population. Our findings underscore the need to consider how structural barriers—such as gendered constraints in negotiation, occupational sorting, or differential access to career-advancing opportunities—may limit the benefits of mobility for highly educated women. Moreover, the observed racial disparities, particularly among Asian graduates, raise important questions about whether returns to mobility are moderated by factors such as occupational segregation, employer biases, or industry-specific mobility patterns. Our findings call for further research into the mechanisms that sustain gender and racial disparities in career advancement.

\section{Abbreviations}
\label{sec:abbreviations}
AI: Artificial intelligence; AMT: Amazon Mechanical Turk; BLS: Bureau of Labor Statistics; BEA: Bureau of Economic Analysis; CIP: Classification of Instructional Programs; ESCO: European Skills, Competences, Qualifications, and Occupations
; FewSOC: Few-shot Prompting Framework for SOC Classification; GDP: Gross Domestic Product; GLMM: Generalized Linear Mixed-Effects Model; GPT: Generating Pre-trained Transformer; HIT: Human Intelligence Task; ICL: In-Context Learning; IRB: Institutional Review Board; ISCO: International Standard Classification of Occupation; LC: Lightcast; LLM: Large Language Model; NAICS: North American Industry Classification System; NLP: Natural Language Processing; O*NET: Occupational Information Network; SOC: Standard Occupational Classification; TF-IDF: Term Frequency-Inverse Document Frequency

\section{Declarations}

\subsection{Ethics approval and consent to participate}
This study was approved by the Institutional Review Board (IRB) of Singapore Management University under Category 1: Exempt from Further IRB Review (approval number IRB-23-142-E039(923)). All participants were recruited through Amazon Mechanical Turk (AMT) and provided informed consent before participating. The study adhered to ethical research guidelines, ensuring voluntary participation and fair compensation. No personally identifiable information was collected.

\subsection{Consent for publication}
All authors have reviewed and approved the manuscript for publication.

\subsection{Clinical trial number}
Not applicable.

\subsection{Availability of data and materials}
All scripts for data cleaning, SOC classification using FewSOC, career trajectory data construction, and regression analyses are publicly available at \url{https://github.com/aekpalakorn/occupation-classification-career-mobility}.

\subsection{Competing interests}
The authors declare that they have no competing interests.

\subsection{Funding}
This research / project is supported by the Ministry of Education, Singapore, under its MOE Academic Research Fund Tier 2 programme (Award T2EP20223-0047). Any opinions, findings and conclusions or recommendations expressed in this material are those of the author(s) and do not reflect the views of the Ministry of Education, Singapore.

\subsection{Authors' contributions}
PA conceptualized the research, developed the methodology, implemented the software, conducted the investigation, performed formal analysis and validation, and was responsible for visualization, writing the original draft, and manuscript review and editing. YX and XJSA contributed to software development and data curation. YL assisted in conceptualization, methodology design, validation, and manuscript review and editing. EPL provided conceptualization, methodology refinement, validation, resources, supervision, project administration, funding acquisition, and manuscript review and editing. All authors read and approved the final manuscript.

\subsection{Acknowledgments}
This research / project is supported by the Ministry of Education, Singapore, under its MOE Academic Research Fund Tier 2 programme (Award T2EP20223-0047). Any opinions, findings and conclusions or recommendations expressed in this material are those of the author(s) and do not reflect the views of the Ministry of Education, Singapore.

\backmatter

\begin{appendices}

\section{Race Classification Evaluation}

The race classification model used for inferring race category from full names in the Career229K dataset was the Logistic Regression model trained on the \textit{Combined} set, an extensive dataset comprising millions of full names from multiple labeled datasets, including the U.S. Census, Wikipedia, and Inmate records. This dataset covers the four primary categories used in the study: White, Black, Asian, and Hispanic. Standard precision (P), recall (R), and F1 scores were used as evaluation metrics.

\begin{table}[htp]
\centering
\caption{Performance of Two-stage ML/Voting Model on the Combined test set}
\label{tab:race_classification}
\begin{tabular}{@{}lccc@{}}
\toprule
\textbf{Racial Category} & \textbf{P} & \textbf{R} & \textbf{F1}\\ \midrule
White & 0.85 & 0.93 & 0.89 \\
Black & 0.72 & 0.47 & 0.57 \\
Asian & 0.75 & 0.4 & 0.52 \\
Hispanic & 0.84 & 0.82 & 0.83 \\ \hline
Macro-average & 0.79 & 0.66 & 0.70 \\ \hline
\end{tabular}
\end{table}

The model achieved a micro-averaged F1 score of 0.83 on the Combined test set, which comprises approximately 1.36 million names. As shown in Table~\ref{tab:race_classification}, performance varies across categories: the model achieved strong F1 scores for White (0.89) and Hispanic (0.83) names but significantly lower scores for Black (0.57) and Asian (0.52) names, resulting in a macro-averaged F1 of 0.70.

This disparity reflects well-documented limitations in name-based classification methods. For the Black category, shared surnames with White Americans mean that classification relies more heavily on distinctive first names; when such patterns are weak or absent, predictive performance declines. The Asian category encompasses diverse subgroups (East Asian, South Asian, and Southeast Asian) with heterogeneous naming conventions, which require broader coverage and diversity in the training data. 

We acknowledge the lower prediction accuracy for Black and Asian names. Nevertheless, this measurement error in the inferred race category variable is expected to impact the main regression analysis in a specific and predictable way. Crucially, measurement error in an independent variable (the inferred race) tends to attenuate estimated coefficients toward zero, rather than introduce spurious associations This attenuation bias (also known as regression dilution bias) is well-documented. Furthermore, systematic misclassification in name-based race inference methods has been observed, with studies indicating that such errors are correlated with demographic and socioeconomic factors \cite{kozlowski2022avoiding}. Taken together, these findings suggest that any observed race-related effects in our regression analyses are likely conservative estimates. Readers should interpret the magnitude of these coefficients with this caution in mind.

\section{Benchmarking Recent LLMs for SOC Classification}

We conducted an additional set of experiments to benchmark the performance of recent, efficient LLMs against GPT-3.5 Turbo for the task of SOC classification. The goal was to assess the current generation of low-latency, low-cost models on the large-scale SOC classification task. The following models were selected:

\begin{itemize}
    \item \textbf{GPT-4.1 Mini}: A recent efficient model from OpenAI, designed for high-throughput classification tasks with low latency and cost.
    \item \textbf{Gemini 2.5 Flash}: The latest small model from Google in the Flash family, designed to balance efficiency and capability. Compared to earlier Flash models, it has enhanced reasoning abilities.
    \item \textbf{Llama 3.1 8B Instruct}: A widely used open-weight model from Meta, representing the current generation of small-to-medium open models.
\end{itemize}

We intentionally excluded large reasoning models such as GPT-5 or Llama 4, as these models incur substantially higher latency and token usage than non-reasoning variants (e.g., GPT-4.1 Mini), making them less practical for large-scale SOC classification and less suitable benchmarks for comparison against GPT-3.5 Turbo.

All models were assessed using the Jobs12K test set and the exact same methodology as the original experiment, including the same FewSOC prompting strategy and evaluation pipeline. All experiments were conducted using the APIs accessible on the OpenAI and Google Vertex AI platforms in October 2025.

While the ground truth in the main paper was established by annotating prediction samples from GPT-3.5 Turbo and the dataset provider (LightCast) using Amazon Mechanical Turk workers, this approach becomes prohibitively expensive and time-consuming when evaluating a large number of models and the expanded pool of predicted labels they generate. 

Therefore, for this experiment, we utilized the LLM-assisted ground truth SOC labels constructed by \citet{achananuparp2025multi}, leveraging the same Jobs12K dataset. This approach utilizes GPT-4o as an expert annotator (LLM-as-a-Judge) to select the final ground truth from a combined pool of predicted candidates. The final ground truth labels represent the majority vote of three independent GPT-4o runs, where the model was tasked with selecting the most appropriate SOC label(s) from a shuffled list of all candidates generated by the tested methods.

\begin{table}[htp]
\centering
\caption{Performance of selected efficient LLMs on the Jobs12K dataset}
\begin{tabular}{@{}ll@{}}
\toprule
\textbf{Model} & \textbf{Accuracy} \\ \midrule
GPT-4.1 Mini & 0.6073 \\
Gemini 2.5 Flash & 0.7425 \\
Llama 3.1 8B Instruct & 0.4722 \\
GPT-3.5 Turbo & 0.7195 \\ \bottomrule
\end{tabular}
\label{tab:llm_benchmark}
\end{table}

The accuracy scores for all models on the Jobs12K test set are presented in Table~\ref{tab:llm_benchmark}. Firstly, Gemini 2.5 Flash achieved the highest accuracy (0.7425), outperforming the GPT-3.5 Turbo benchmark by 2.3 percentage points. Despite being a newer efficient GPT-based model, GPT-4.1 Mini (0.6073) performed significantly poorer than GPT-3.5 Turbo (0.7195). This unexpected result suggests that the architectural trade-offs made to optimize GPT-4.1 Mini may have compromised the model's ability to handle SOC classification. Lastly, The Llama 3.1 8B Instruct model lagged substantially behind all the other models, achieving an accuracy of 0.4722. This finding is consistent with recent studies \cite{achananuparp2025multi}, which similarly showed the comparatively poor performance of Llama models in the SOC classification task.

\section{Multilevel Sensitivity Analysis}

We conducted a multilevel sensitivity analysis using Generalized Linear Mixed Models (GLMMs) to test the robustness of the main effects estimated in our primary logistic regression models (Models 1–4) against unobserved within-cluster heterogeneity. Specifically, we fit several mixed-effect logistic regression models using the Career229K dataset and Binomial Bayes Mixed GLMs in StatsModels v.0.14.

The goals of the sensitivity analysis were two-fold: (1) To verify the stability of the direction and significance of main effects and interactions identified in Models 1–4 when accounting for unobserved cluster-level variance; and (2) to estimate more precise cluster-specific fixed effects, controlling for heterogeneity at higher levels such as major occupations, major industries, states, etc.

\subsection{Random Effects Structure}

We systematically incorporated five sources of clustering, summarized in Table~\ref{tab:random_effects}, including Major Occupation, Major Industry, State, Job Start Year (Career Entry) Cohort, and Firm Size Group. Each random effect represents a potential source of unobserved heterogeneity in upward mobility outcomes.

\begin{table}[!ht]
\centering
\caption{Random Effect Components and Descriptions}
\begin{tabular}{@{}lll@{}}
\toprule
\textbf{Label} & \textbf{Random Effect} & \textbf{Description} \\ 
\midrule
A & Major Occupation & \makecell[l]{Major O*NET occupation (2-digit SOC code)} \\
B & Major Industry & \makecell[l]{Major NAICS industry (2-digit NAICS code)} \\
C & State & \makecell[l]{US State} \\
D & Job Start Year & \makecell[l]{Career entry year cohort} \\
E & Firm Size Group & \makecell[l]{Firm size group (Mega, Large, Medium, Sparse)} \\
\bottomrule
\end{tabular}
\footnotetext{All random effects are measured at the start of the career.}
\label{tab:random_effects}
\end{table}

State and Job Start Year were included to account for regional and cohort clustering effects, respectively. Although these variables weren't utilized in the main paper's analysis, they are available in the Career229K dataset and allow us to model location- and career-entry-specific heterogeneity in upward mobility outcomes.

Since most individual firms have very few observations (i.e., individual worker profiles), using individual firms as a random effect would produce unstable estimates. To address this, we introduced a Firm Size Group variable, aggregating firms into four categories based on the number of workers represented in the dataset: Mega ($>$300 workers), Large (101–300 workers), Medium (10–100 workers), and Sparse ($<$10 workers). These thresholds were heuristically selected to balance granularity with sufficient sample sizes per group, ensuring stable random effect estimates while capturing meaningful organizational differences. Mega firms generally correspond to large corporations in sectors such as retail, technology, finance, and consulting (e.g., Target, Google, Meta), whereas Sparse firms include small or niche organizations. By modeling firm size groups rather than individual firms, we capture firm-level heterogeneity without introducing sparsity-related instability.

\subsection{Robustness Tests for Individual Random Effects}

We first evaluated the influence of each clustering factor independently by estimating Model 1 (main effects only) under five specifications, denoted Models 1A through 1E:

\begin{itemize}
\item Model 1A: Random intercept for Major Occupation
\item Model 1B: Random intercept for Major Industry
\item Model 1C: Random intercept for State
\item Model 1D: Random intercept for Job Start Year
\item Model 1E: Random intercept for Firm Size Group
\end{itemize}

As shown in Table~\ref{tab:models_1A_1E}, the direction and credibility of the main fixed effects -- gender, race, educational attainment, and generational cohort -- remained consistently stable across all model specifications, with most effects showing 95\% credible intervals excluding zero (``credible effects''). Random-effect variance components suggest that all five clusterings contribute meaningfully to unexplained heterogeneity, with particularly notable variance attributable to Firm Size Group and Major Occupation.

\begin{table}[!ht]
\centering
\caption{Posterior means for fixed effects, interaction terms, and random effects in Models 1A--1E}
\begin{tabular}{@{}lccccc@{}}
\toprule
\textbf{Variable} & \textbf{1A} & \textbf{1B} & \textbf{1C} & \textbf{1D} & \textbf{1E} \\ 
\midrule
Intercept & -1.1047* & -1.1046* & -0.9937* & -1.3167* & -1.0581* \\
Female & -0.1534* & -0.1534* & -0.1566* & -0.1546* & -0.1520* \\
Black & -0.0703* & -0.0703* & -0.0690* & -0.0730* & -0.0717* \\
Asian & 0.1331* & 0.1331* & 0.1302* & 0.1288* & 0.1270* \\
Hispanic & 0.0234 & 0.0234 & 0.0217 & 0.0192 & 0.0236 \\
Doctorate & 1.2081* & 1.2081* & 0.3117* & 0.0975* & 0.0135* \\
Master's Degree & 0.4090* & 0.4090* & -0.0045* & 0.1792* & 0.0516* \\
Millennials & 0.1744* & 0.1744* & 0.0416* & 0.2674* & -0.0745* \\
Regional Economic Ranking & 0.0416* & 0.0416* & -0.1984* & 0.3130* & 1.2010* \\
Log(Wage) & -0.9984* & -0.9984* & 0.0672* & 0.3298* & 0.1797* \\
Job Mobility & 0.5950* & 0.5950* & 0.0105* & 0.3360* & 0.4053* \\
Type-1 Change & 0.9354* & 0.9354* & 0.9333* & 0.9316* & 0.9409* \\
Type-2 Change & 0.9527* & 0.9527* & 0.9529* & 0.9492* & 0.9483* \\
Type-3 Change & 0.7982* & 0.7982* & 0.7991* & 0.7974* & 0.8054* \\
\midrule
Major Occupation (RE) & 0.0138* & -- & -- & -- & -- \\
Major Industry (RE) & -- & 0.0131* & -- & -- & -- \\
State (RE) & -- & -- & 0.0122* & -- & -- \\
Job Start Year (RE) & -- & -- & -- & 0.0127* & -- \\
Firm Size Group (RE) & -- & -- & -- & -- & 0.0365* \\
\bottomrule
\end{tabular}
\label{tab:models_1A_1E}
\footnotetext{Note: \textit{Reference Categories:} Male (Gender), White (Race), Bachelor’s degree (Educational Attainment), Generation X (Social Generation), Management (Occupation), and Agriculture, Forestry, Fishing, and Hunting (Industry). Posterior means of fixed-effect coefficients and random-effect (RE) variance components are shown. * denotes effects whose 95\% credible intervals exclude zero.}
\end{table}

\subsection{Joint Random Effects Structure and Interaction Models}

Based on the posterior variance estimates from individual random-effect components, we identified Firm Size Group and Major Occupation as the most influential sources of unobserved heterogeneity in upward mobility outcomes. These clusters correspond to meaningful organizational and occupation-specific contexts that are theoretically expected to predict career trajectories. For instance, larger firms may offer more structured mobility pathways and resources, while occupations differ in skill requirements, labor market demand, and advancement opportunities.

Accordingly, we constructed a more stringent random-effects structure for the subsequent interaction models, jointly including these two components as random intercepts while retaining all group-level variables as fixed effects. This specification allows us to account for the largest sources of baseline heterogeneity without overly increasing model complexity, ensuring stable and interpretable estimates of population-averaged interaction effects.

The resulting mixed-effect model specifications were as follows:

\begin{itemize}
\item Model 1F: Main effects with random intercepts for Major Occupation and Firm Size Group
\item Model 2F: Main effects and Gender × Job Change Type interactions with the same random intercepts
\item Model 3F: Main effects and Race × Job Change Type interactions with the same random intercepts
\item Model 4F: Main effects and Race × Job Change Type interactions stratified by gender with the same random intercepts
\end{itemize}

These models allow us to test interaction hypotheses while controlling for the most critical cluster-level heterogeneity.

Tables~\ref{tab:models_1F_4F} presents the posterior means of fixed-effect coefficients and random-effect variance components for Models 1F–4F. Asterisks (*) indicate effects whose 95\% credible intervals exclude zero.

\subsection{Results and Discussion}

\begin{figure}[ht]
\centering
\includegraphics[width=0.9\textwidth]{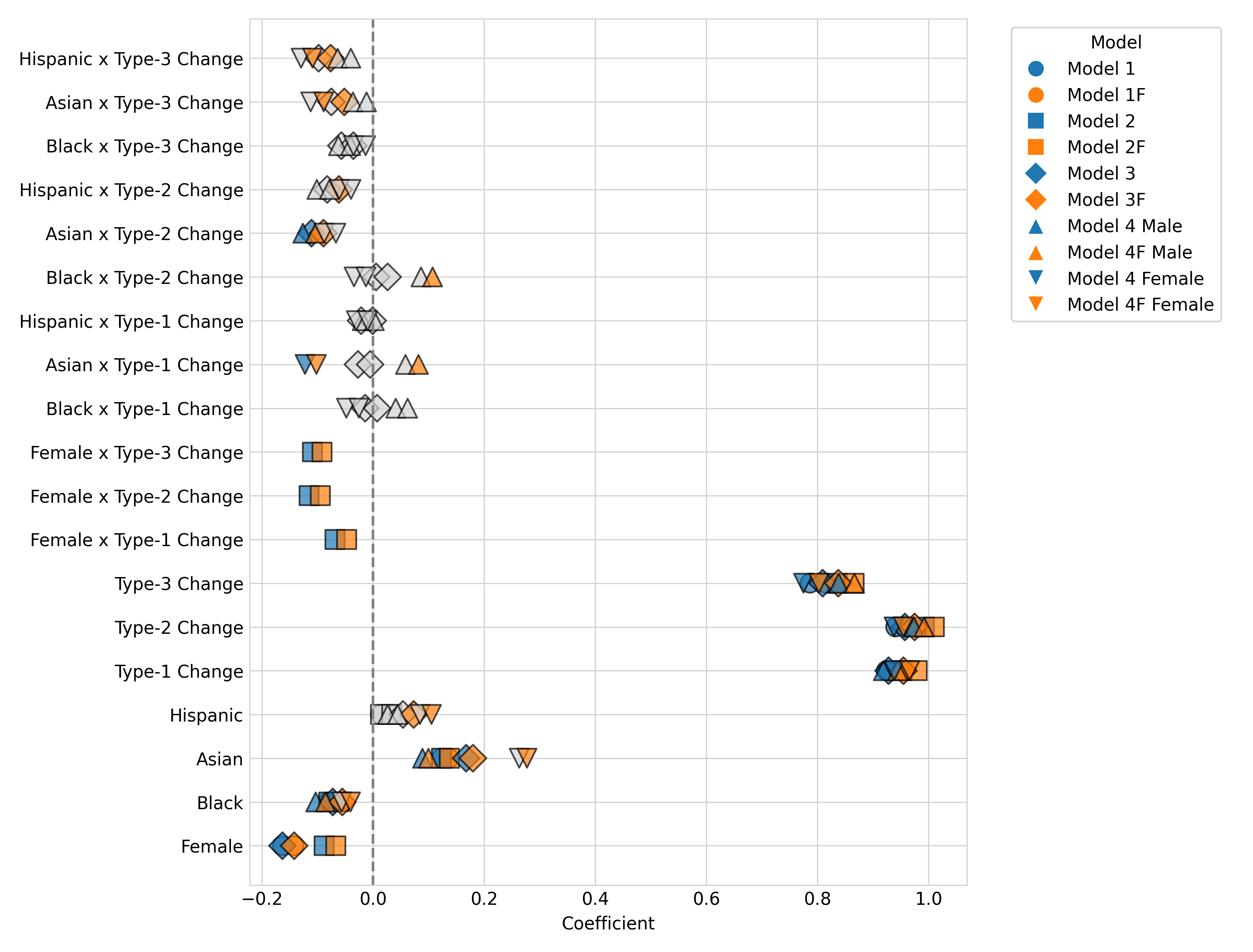}
\caption{Coefficient plot of estimated effects for main predictors and key interaction terms across Models 1–4 and 1F–4F. Statistical significance and credibility are indicated through color: coefficients significant at ($p<0.05$) or deemed credible are shown in distinct model-specific colors, whereas non-significant ($p>0.05$) or non-credible estimates are rendered in gray. To enhance visual discrimination when estimates are similar, markers are horizontally jittered slightly.}
\label{fig:sensitivity_main_effects_plot}
\end{figure}

Table \ref{tab:models_1F_4F} reports the posterior means of fixed-effect coefficients, interaction terms, and random-effect variance components for Models 1F–4F. To complement the tabular presentation and aid comparison of estimates across models, Figures \ref{fig:sensitivity_main_effects_plot} and \ref{fig:sensitivity_covariates_plot} provide coefficient plots that visualize the same estimates along side the corresponding main effects from Models 1 -- 4. Statistical significance and credibility are visually encoded through color: coefficients significant at ($p<0.05$) or whose 95\% credible intervals exclude zero are displayed in distinct model-specific colors, whereas non-significant or non-credible estimates are rendered in gray. To enhance visual discrimination when estimates are similar in magnitude, markers are slightly jittered horizontally. Figure \ref{fig:sensitivity_main_effects_plot} focuses on the main predictors and key interaction terms, while Figure \ref{fig:sensitivity_covariates_plot} presents the full set of control variables.

\begin{table}[htbp]
\centering
\caption{Posterior means for fixed effects, interaction terms, and random effects in Models 1F–4F}
\begin{tabular}{@{}lcccccc@{}}
\toprule
\multirow{2}{*}{Variable} & \multicolumn{1}{c}{1F} & \multicolumn{1}{c}{2F} & \multicolumn{1}{c}{3F} & \multicolumn{2}{c}{4F} \\ \cmidrule(l){5-6}
 &  &  &  & Male & Female \\ \midrule
Intercept & -1.058$^{*}$ & -1.0958$^{*}$ & -1.0669$^{*}$ & -1.2127$^{*}$ & -1.0501$^{*}$ \\
Female & -0.152$^{*}$ & -0.077$^{*}$ & -0.1518$^{*}$ & - & - \\
Black & -0.0717$^{*}$ & -0.0718$^{*}$ & -0.0652$^{*}$ & -0.0955$^{*}$ & -0.0504$^{*}$ \\
Asian & 0.127$^{*}$ & 0.1274$^{*}$ & 0.1699$^{*}$ & 0.0898$^{*}$ & 0.2672$^{*}$ \\
Hispanic & 0.0236 & 0.0238 & 0.0632$^{*}$ & 0.0343 & 0.0953$^{*}$ \\
Doctorate & 1.201$^{*}$ & 1.2017$^{*}$ & 1.2005$^{*}$ & 1.1791$^{*}$ & 1.2514$^{*}$ \\
Master's Degree & 0.4053$^{*}$ & 0.4054$^{*}$ & 0.405$^{*}$ & 0.4017$^{*}$ & 0.413$^{*}$ \\
Millennials & 0.1798$^{*}$ & 0.1796$^{*}$ & 0.1794$^{*}$ & 0.1891$^{*}$ & 0.1689$^{*}$ \\
Regional Economic Ranking & 0.0364$^{*}$ & 0.0365$^{*}$ & 0.0364$^{*}$ & 0.0235$^{*}$ & 0.0511$^{*}$ \\
Log(Wage) & -1.0004$^{*}$ & -1.0003$^{*}$ & -1.0006$^{*}$ & -0.9622$^{*}$ & -1.0337$^{*}$ \\
Job Mobility & 0.5935$^{*}$ & 0.5936$^{*}$ & 0.5933$^{*}$ & 0.6058$^{*}$ & 0.5804$^{*}$ \\
Type-1 Change & 0.9409$^{*}$ & 0.9692$^{*}$ & 0.9446$^{*}$ & 0.9343$^{*}$ & 0.9558$^{*}$ \\
Type-2 Change & 0.9484$^{*}$ & 1.0005$^{*}$ & 0.9647$^{*}$ & 0.9819$^{*}$ & 0.9445$^{*}$ \\
Type-3 Change & 0.8054$^{*}$ & 0.856$^{*}$ & 0.8274$^{*}$ & 0.8564$^{*}$ & 0.7925$^{*}$ \\
Female $\times$ Type-1 Change & - & -0.0573$^{*}$ & - & - & - \\
Female $\times$ Type-2 Change & - & -0.1049$^{*}$ & - & - & - \\
Female $\times$ Type-3 Change & - & -0.1015$^{*}$ & - & - & - \\
Black $\times$ Type-1 Change & - & - & -0.0026 & 0.0524 & -0.0354 \\
Asian $\times$ Type-1 Change & - & - & -0.0152 & 0.0719$^{*}$ & -0.1118$^{*}$ \\
Hispanic $\times$ Type-1 Change & - & - & -0.0105 & -0.0076 & -0.0197 \\
Black $\times$ Type-2 Change & - & - & 0.0164 & 0.0973$^{*}$ & -0.0226 \\
Asian $\times$ Type-2 Change & - & - & -0.0991$^{*}$ & -0.1135$^{*}$ & -0.0767 \\
Hispanic $\times$ Type-2 Change & - & - & -0.0713$^{*}$ & -0.0885 & -0.0496 \\
Black $\times$ Type-3 Change & - & - & -0.0453 & -0.05 & -0.0232 \\
Asian $\times$ Type-3 Change & - & - & -0.0616$^{*}$ & -0.0216 & -0.0987$^{*}$ \\
Hispanic $\times$ Type-3 Change & - & - & -0.0862$^{*}$ & -0.0498 & -0.1187$^{*}$ \\ \midrule
Firm Size Group (RE) & 0.0365$^{*}$ & 0.0365$^{*}$ & 0.0365$^{*}$ & 0.0425$^{*}$ & 0.0436$^{*}$ \\
Major Occupation (RE) & 0.0138$^{*}$ & 0.0138$^{*}$ & 0.0138$^{*}$ & 0.0186$^{*}$ & 0.0186$^{*}$ \\ \bottomrule
\end{tabular}
\label{tab:models_1F_4F}
\footnotetext{Note: \textit{Reference Categories:} Male (Gender), White (Race), Bachelor’s degree (Educational Attainment), Generation X (Social Generation), Management (Occupation), and Agriculture, Forestry, Fishing, and Hunting (Industry). Posterior means of fixed-effect coefficients and random-effect (RE) variance components are shown. * denotes effects whose 95\% credible intervals exclude zero.}
\end{table}

\textbf{Main Effects (Model 1F)}:
Models 1F–4F confirm that the primary main-effect findings from Models 1–4 are robust to the inclusion of the most influential random effects. Posterior means for Female, Black, and Asian remain credible and directionally consistent across all models, indicating that gender and racial disparities in upward mobility are not explained away by cluster-level heterogeneity in occupation, firm size, or other sources.

In contrast, the Hispanic effect, which was not significant across all main models, becomes credible in two of the interaction specifications -- Models 3F and 4F for female workers. This suggests that once unobserved heterogeneity is appropriately modeled, previously masked racial advantages for Hispanic workers emerge in specific contexts, particularly among Hispanic women in certain job-mobility patterns. This nuance refines RQ2’s findings by revealing that Hispanic effects are not uniformly weak but rather contingent on mobility type and intersectional factors, which were not fully captured in the simpler models.

\textbf{Gender × Job Change Type Interactions (Model 2F)}:
Introducing gender-by-movement type interactions reveals that women face small but credible disadvantages across all three types of job change, compared to men. The coefficients for Female × Type-1, Female × Type-2, and Female × Type-3 changes are all negative and credible, consistent with Model 2. Importantly, these effects persist after accounting for random effects at the occupation and firm-size levels, indicating that gendered disparities in upward mobility are not fully attributable to differences in where women and men work.

\textbf{Race × Job Change Type Interactions (Model 3F)}:
The inclusion of race-by-movement type interactions uncovers additional heterogeneity not visible in the baseline model. For example, a baseline Hispanic advantage emerges in Type-2 and Type-3 job changes, which was masked when interaction terms were omitted. Conversely, Asian workers experience additional credible disadvantages in Type-3 job change, which was absent from the baseline model.

\textbf{Stratified Race × Job Change Type Interactions (Model 4F)}:
Stratifying by gender reveals additional intersectional patterns in the relationship between race and career mobility that were not visible in the aggregate models.

For men, the negative Asian × Type-2 effect observed in Model 3F remains robust, indicating persistent disadvantages for Asian men in intra-firm occupation changes. In addition, Model 4F uncovers new credible interaction effects, including positive \emph{Asian × Type-1} and positive Black × Type-2, indicating complex dynamics between upward mobility, race, and job change types among male workers.

For women, the negative Asian × Type-1 effect persists, confirming a robust disadvantage for Asian women in inter-firm occupation changes. Model 4F also reveals new credible disadvantages for Asian × Type-3 and Hispanic × Type-3, highlighting intersectional barriers for Asian and Hispanic women in inter-firm lateral moves.

\textbf{Random Effects Contributions}:
Across all models, the inclusion of random effects does not fundamentally change the fixed-effect estimates but provides insights into the sources of residual heterogeneity. Variance components associated with Major Occupation, Major Industry, State, Job Start Year, and Firm Size Group indicate that these clustering factors contribute meaningfully to unexplained variation in upward mobility, but they do not eliminate or reverse observed disparities, confirming the robustness of Models 1–4.

\section{Supplementary Figures}

\begin{figure}[ht]
\centering
\includegraphics[width=0.9\textwidth]{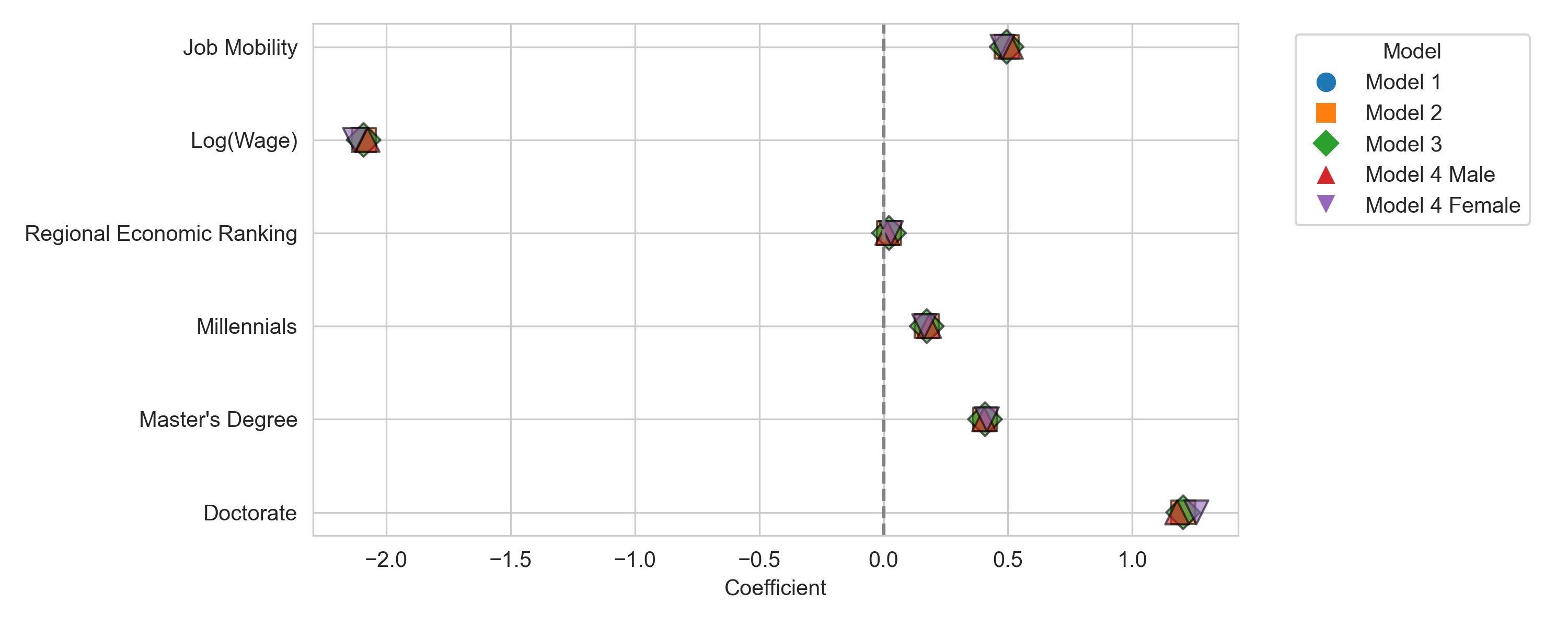}
\caption{Coefficient plot of estimated coefficients for covariates in Models 1--4. Statistical significance is visually encoded: Coefficients significant at $p<0.05$ are displayed in their full, distinct model colors, while insignificant coefficients ($p>0.05$) are displayed using gray-filled shapes.}
\label{fig:covariates_plot}
\end{figure}

\begin{figure}[ht]
\centering
\includegraphics[width=0.9\textwidth]{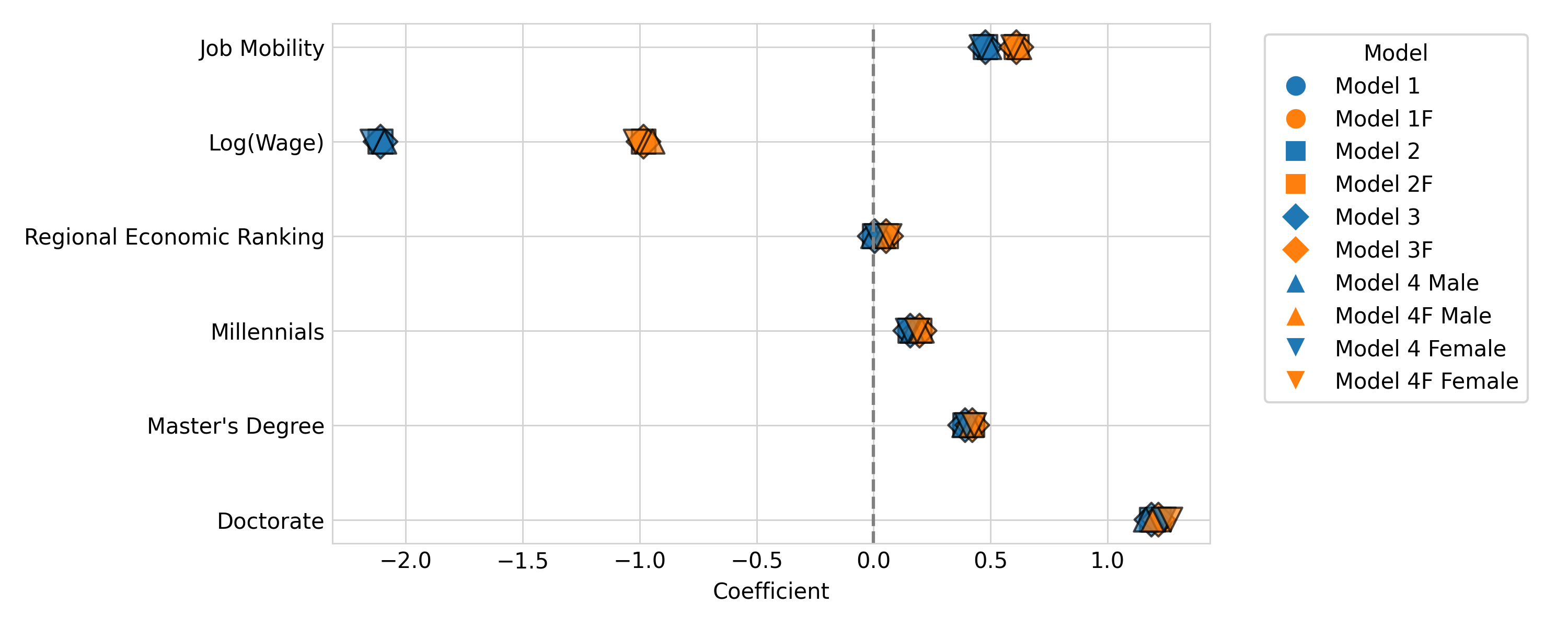}
\caption{Coefficient plot of estimated effects for covariates across Models 1–4 and 1F–4F. Statistical significance and credibility are indicated through color: coefficients significant at ($p<0.05$) or deemed credible are shown in distinct model-specific colors, whereas non-significant ($p>0.05$) or non-credible estimates are rendered in gray. To enhance visual discrimination when estimates are similar, markers are horizontally jittered slightly.}
\label{fig:sensitivity_covariates_plot}
\end{figure}

Figures~\ref{fig:covariates_plot} and~\ref{fig:sensitivity_covariates_plot} display coefficient plots for all covariates corresponding to Tables~\ref{tab:regression_results} and~\ref{tab:models_1F_4F}, respectively. The plots highlight the stability of the control variables, with coefficient estimates tightly clustered across all model specifications.

\section{Prompt Template and Examples}
\label{sec:prompt_template}

\begin{tcolorbox}[colback=blue!5!white, colframe=blue!75!black, title=SOC Generation, label=prompt:soc_generation]
\small
You are an expert O*NET-SOC 2019 coder. Given a list of job title-company name pairs in input texts, assign the following labels for each input text:\\
1. Occupational title and code from the O*NET-SOC 2019. Separate the title and code with colon. If no suitable answer is available, a best or random guess is fine. If the input text mentions a student, answer `Student'.\\
2. Yes (Y) or no (N) to whether the input text mentions a non-occupational role or not. Non-occupational roles typically include keywords such as intern, student, volunteer, founder, owner, member, etc.\\
3. Yes (Y) or no (N) to whether the input text mentions multiple roles or not.\\
Separated each label with semi-colon. Do not explain.\\
---\\
Answer format:\\
Task ID; Label 1; Label 2; Label 3\\
---\\
Examples:\\
$\{\{examples\}\}$\\
---\\
Input texts:\\
$\{\{data\}\}$\\
Answers:
\end{tcolorbox}

\scalebox{0.9}{
\begin{tcolorbox}[colback=blue!5!white, colframe=blue!75!black, title=SOC Generation: Few-shot Examples, label=prompt:examples]
\small
Input texts:\\
T1; Aircraft cabin cleaner, Avionic Services\\
T2; Barista, Starbucks Coffee\\
T3; Stage manager, director, and owner, Old Vic theatre\\
...\\
T17; Summer research intern, Columbia University\\
Answers:\\
T1; Aircraft Service Attendants:53-6032.00; N; N\\
T2; Baristas:35-3023.01; N; N\\
T3; Producers and Directors:27-2012.00; Y; Y\\
...\\
T17; Social Science Research Assistants:19-4061.00; Y; N
\end{tcolorbox}}

\begin{tcolorbox}[colback=blue!5!white, colframe=blue!75!black, title=SOC Selection, label=prompt:soc_selection]
\small
Select the O*NET-SOC that best describes each job title-company name pair below. Choose only one number from the options provided. Do not explain.\\
---\\
Examples:\\
T1. social media manager, mcs - midwest conference service // options: 1. Public Relations Managers (11-2032.00); 2. Fundraising Managers (11-2033.00)\\
T2. medical scribe, proscribe // options: 1. Medical Records Specialists (29-2072.00); 2. Health Information Technologists and Medical Registrars (29-9021.00)\\
Answers:\\
T1:1\\
T2:1\\
---\\
\{\{data\}\}\\
Answers:
\end{tcolorbox}
\end{appendices}

\bibliography{main}

\end{document}